\documentclass[fleqn,usenatbib,onecolumn]{mnras}

\usepackage{newtxtext,newtxmath}
\usepackage[T1]{fontenc}

\DeclareRobustCommand{\VAN}[3]{#2}
\let\VANthebibliography\thebibliography
\def\thebibliography{\DeclareRobustCommand{\VAN}[3]{##3}\VANthebibliography}
\usepackage{graphicx,xcolor}	% Including figure files
\usepackage{amsmath}	% Advanced maths commands

\graphicspath{{./}{figures/}}
\newcommand{\psj}{Planet. Sci. J.}
\title{Milankovitch Cycles for a Circumstellar Earth-analog within $\alpha$ Centauri-like Binaries}

\author[B. Quarles et al.]{
B. Quarles,$^{1,2}$\thanks{E-mail: billylquarles@gmail.com}
G. Li,$^{1}$
and J.~J. Lissauer$^{3}$
\\
$^{1}$Center for Relativistic Astrophysics, School of Physics, 
Georgia Institute of Technology, Atlanta, GA 30332, USA\\
$^{2}$Department of Physics, Astronomy, Geosciences and Engineering Technology, Valdosta State University, Valdosta GA, 31698, USA\\
$^{3}$Space Science and Astrobiology Division,
MS 245-3, NASA Ames Research Center, Moffett Field, CA 94035, USA
}

\date{Accepted XXX. Received YYY; in original form ZZZ}

\pubyear{2021}

\begin{document}
\label{firstpage}
\pagerange{\pageref{firstpage}--\pageref{lastpage}}
\maketitle

\begin{abstract}
An Earth-analog orbiting within the habitable zone of $\alpha$ Centauri B was shown to undergo large variations in its obliquity, or axial tilt, which affects the planetary climate by altering the radiative flux for a given latitude.  We examine the potential implications of these obliquity variations for climate through Milankovitch cycles using an energy balance model with ice {growth and retreat}.  Similar to previous studies, the largest amplitude obliquity variations from spin-orbit resonances induce snowball states within the habitable zone, while moderate variations can allow for persistent ice caps or an ice belt.  Particular outcomes for the global ice distribution can depend on the planetary orbit, obliquity, spin precession, binary orbit, and which star the Earth-analog orbits.  An Earth-analog with an inclined {orbit relative to the binary orbital plane} can periodically transition through several global ice distribution states and risk runaway glaciation when {ice appears at both poles and the equator}.  When determining the potential habitability for planets in {general} stellar binaries, more care must be taken due to the orbital and spin dynamics.  {For Earth-analogs within the habitable zone of $\alpha$ Centauri B can experience a much greater range of climate states, which is in contrast to Earth-analogs in the habitable zone of $\alpha$ Centauri A.}
\end{abstract}


%%%%%%%%%%%%%%%%%%%%%%%%%%%%%%%%%%%%%%%%%%%%%%%%%%

%%%%%%%%%%%%%%%%% BODY OF PAPER %%%%%%%%%%%%%%%%%%

\section{Introduction} \label{sec:intro}
Current surveys of stellar multiplicity indicate that nearly half of Sunlike stars have stellar companions \citep{Raghavan2010,Moe2017}.  The most recent census of Kepler planets \citep{Bryson2021} suggests that nearly half of all Sunlike stars harbor an Earthlike planet in its habitable zone (i.e., orbital region where liquid water could potentially exist).  As a result, the prospect of planets in binary systems is now more compelling. New observations using direct imaging have uncovered a possible Neptune-sized planet orbiting $\alpha$ Centauri A \citep{Wagner2021}, which is the primary star within the closest stellar system to the Sun ($\sim4$ ly).  Future observations using the James Webb Space Telescope (JWST) will help identify the large planets orbiting $\alpha$ Centauri A \citep{Beichman2020}, which will also have implications for prospective planets orbiting the companion star, $\alpha$ Centauri B.  The habitability of planets orbiting either star depends on more factors than those typically used for single star systems \citep{Kasting1993}.

The habitable zone (HZ), as defined for our solar system, is typically calculated using atmospheric models with a single energy source \citep{Kopparapu2013,Kopparapu2013b,Kopparapu2014}.  The HZ definition has expanded to include binary stars {\citep{Eggl2012,Kaltenegger2013,Haghighipour2013,Cukier2019,Eggl2020}}, where the spectral energy distribution of each star must be taken into account when the radiative flux from the secondary star is non-negligible.  \cite{Forgan2012} showed the extent of the HZ around $\alpha$ Centuari B is largely constant, despite the radiative influence from $\alpha$ Centuari A.  However, planets within either star's HZ would undergo eccentricity \citep{Quarles2016,Quarles2018a} and obliquity oscillations \citep{Quarles2019} that can affect the long-term climate stability even when the overall extent of the HZ is static.  Perturbations on Earth's orbit and spin precession induce so-called Milankovitch cycles \citep{Milankovitch1969} that modulate the extent and frequency of ice ages.  Large scale changes in the extent of ice {coverage} on Earth's surface influence the climate energy balance through albedo and associated feedbacks \citep[i.e.,][]{Spiegel2010,Armstrong2014,Deitrick2018}.

\cite{Forgan2016} coupled a 1D latitudinal energy balance model (LEBM) to N-body integrations and showed that perturbations from the stellar binary drive variations in the mean temperature for an Earth-like planet.  For simulations of an Earth-like planet orbiting $\alpha$ Centauri A, the magnitude of variations was $\sim$5 K over a range of timescales depending on whether the types of precession (orbital or axial) considered.  \cite{Forgan2016} focused on two archetypal systems, but showed that the decoupled approach of considering the astronomical (i.e., orbit and spin) and radiative perturbations separately was broadly acceptable.  Later works \citep{Bazso2017,Quarles2018a} showed conditions where the maximum planetary eccentricity is non-negligible ($e_{\rm p}\sim$0.15) due to the secular forcing.  Moreover, \cite{Quarles2019} illustrated large variations in the planetary obliquity, which have broader implications concerning the periodicity of ice ages.  Therefore, we revisit the study of Milankovitch cycles for circumstellar planets in binaries using $\alpha$ Centauri as an archetype.

%Paragraph about what science we will address
This work focuses the influence of stellar companions on the growth and retreat of ice {coverage} as commonly associated with Milankovitch cycles, which can also be used as a proxy for habitability.  Thus, we use a 1D LEBM similar to \cite{Forgan2016} and forego the more detailed approach employed through a general circulation model (GCM).  The $\alpha$ Centauri system is included in our investigation because the two most massive stars form a binary consisting of Sun-like star of disparate luminosity, which allows for a more rich exploration.  \cite{Quarles2019} showed conditions that produce large obliquity variations for an Earth-analog within $\alpha$ Cen B's HZ, while other works \citep{Armstrong2014,Deitrick2018,Shan2018,Quarles2020a} have investigated the capture of planets into a Cassini state (i.e., a 1:1 resonance between the spin angle $\psi$ and ascending node $\Omega$; \cite{Colombo1966,Peale1969,Ward2004}), where such processes can directly affect the extent of Milankovitch cycles.

%Summary paragraph outlining the sections of the paper
This paper explores the effects of obliquity variation on the potential for Milankovitch cycles when an Earth-analog orbits one star within a stellar binary.  Section \ref{sec:num_meth} details our initial conditions and physical processes implemented in the numerical simulations.  The influence of obliquity variation on the globally averaged surface temperature, albedo, ice fraction, and ice distribution are discussed in Sections \ref{sec:aCenA} and \ref{sec:aCenB} for an Earth-analog orbiting either star in $\alpha$ Centauri AB.  A broader study is performed in Section \ref{sec:starB} for Earth-like planets orbiting the secondary star of a binary that is $\alpha$ Centuari-like with respect to the stellar mass and luminosity, but the binary orbit varies over a range in binary semimajor axis and eccentricity.  The main results our our study are summarized in Section \ref{sec:conc}, along with a comparison to previous results.

% The possible observable signatures are discussed in Section \ref{sec:obs}.  

\section{Numerical Methods} \label{sec:num_meth}
\subsection{Orbital Evolution} \label{sec:orbit}
% Long-term simulations of orbits for planets in binary systems typically requires a specialized numerical scheme \citep[e.g.,][]{Chambers2002} that evolves the binary orbit separately from the planetary orbits for efficiency over billion year timescales.  

Many studies \citep{Wiegert1997,Holman1999,Quarles2016,Quarles2018a,Quarles2018b,Quarles2020b} investigated the stability of planets in binaries, including $\alpha$ Centauri AB, and concluded that planetary orbits within the HZ are generally stable, particularly those with low eccentricity ($\lesssim0.2$) and inclination ($\lesssim40^\circ$) relative to the binary orbit.  As a result, we use the \texttt{whfast} integrator with \texttt{REBOUND} \citep{Rein2012,Rein2015} to evolve each system.  The timestep for the integrator is set to 5\% of the planetary period, which is adequate to keep the numerical errors and simulation wall time low.     

Observations of $\alpha$ Centauri AB have improved over time, and we use parameters from \citet{Pourbaix2016}.  The stellar masses are 1.133 M$_\odot$ (star A) and 0.972 M$_\odot$ (star B), where the orbital parameters are 23.78 au, 0.524, 77.05$^\circ$ and 209.6901$^\circ$ for the initial binary semimajor axis $a_{\rm bin}$, eccentricity $e_{\rm bin}$, argument of periastron $\omega_{\rm bin}$, and mean anomaly $MA_{\rm bin}$, respectively.  The stellar luminosities are 1.519 L$_\odot$ (star A) and 0.5 L$_\odot$ (star B), which will be used to scale the planetary orbits within the climate model.  {For Earth-analogs orbiting either star in $\alpha$ Cen AB, the} planetary orbit begins with a semimajor axis $a_{\rm p}$ (in AU) so that the planet receives an Earth-equivalent amount of radiative flux {$S(=S_\oplus)$} at the top of its atmosphere through the relation, $a_{\rm p} = \sqrt{L/S}$, {using the luminosity in solar units $L_\odot$}.  The stellar pericenter distance is $\sim$11 au and the radiative flux for a planet orbiting star B can increase up to 1.3\% at conjunction, while the increase is  smaller (0.5\%) for an Earth-analog orbiting star A \citep{Quarles2016}.  The flux contribution from the stellar companion near its pericenter in either case is significant only at conjunctions and the magnitude of the contribution is much less than the increased flux from the host star at the planet's pericenter.  Therefore, we ignore the flux contribution of the stellar companion and focus on its gravitational influence.   Additionally, the planetary orbit is apsidally aligned with the binary (i.e., $\omega_{\rm p}=\omega_{\rm bin}$) and near the forced eccentricity (see \cite{Quarles2018a}).  {We note that there are stable orbital solutions without our condition of apsidal alignment that result in eccentricity oscillations.  We focus on the apsidally aligned case, where the obliquity variations will dominate over the eccentricity variations in the climate model,  to reduce the complexity of results.  Although, recent planet formation models in binary systems \cite[e.g.,][]{Martin2020,Silsbee2021} showed that disks with apsidal alignment favor conditions for successful planetesimal growth.}  {For planetary orbits around either star, the} initial planetary mean anomaly is 222.492236$^\circ$. All of our results will be averaged over a planetary orbit, which removes any dependence of the planetary mean anomaly. Our previous work \citep{Quarles2019} explored a range of mutual planetary inclinations in detail, where we restrict this study to include planetary orbits tilted by 2$^\circ$, 10$^\circ$, and 30$^\circ$.

Most of our simulations are evaluated for 0.5 Myr, which captures many secular cycles of the planetary orbit in response to the binary companion.  But, we also perform a second set of simulations studying general stellar binaries beyond $\alpha$ Centauri AB. We set the binary semimajor axis ranges from 10--90 au (in 1 au steps) and eccentricity ranges from 0.0--0.9 (in 0.01 steps).  In these simulations, the planet orbits star B with an inclination of 10$^\circ$.  Due to spin-orbit resonances the timescale for planetary obliquity oscillations can reach 1 Myr and thus, we evolve these cases up to 2 Myr to account for the longer secular timescale.  

\subsection{Obliquity Evolution} \label{sec:obliq}
The torque of the host star on the planet's quadrupole moment induces a spin precession on the Earth-analog, which we account for using the parameter $\gamma$\footnote{This parameter is changed from the conventional $\alpha$ for the precession constant to clearly distinguish with the $\alpha$ used for albedo in the climate model.} in arcseconds/yr.  Values for $\gamma$ are proportional to the square of the planetary rotation rate and the planetary $J_2$.  Previous works \citep{Barnes2016,Quarles2020a} used an algorithm from \cite{Lissauer2012} to calculate the planetary $J_2$ from the planetary rotation period.  We use a similar procedure and produce a table of $J_2$ values and planetary periods that correspond to our prescribed $\gamma$.  For each star in $\alpha$ Cen AB, we evaluate a grid of initial spin states with $\gamma$ ranging from 0-100\arcsec/yr and prograde planetary obliquity $\varepsilon_o$ from 0$^\circ$--90$^\circ$.  {We focus our study on prograde rotation because our previous studies \citep{Barnes2016,Quarles2019,Quarles2020a} showed that the positive spin precession frequencies for retrograde are very low amplitude and require a moderate--large planetary eccentricity ($e_{\rm p} \gtrsim 0.3$) to become relevant \citep{Kreyche2020}.  Some initial obliquities ($\varepsilon_o \sim 89^\circ$) can evolve into the retrograde regime ($\varepsilon>90^\circ$), for a limited time, where the maximum obliquity is less than $93^\circ$.}  The planetary spin vector is defined by two angles: obliquity $\varepsilon$ and spin longitude $\psi$, where \cite{Quarles2019} explored how the resulting obliquity variation $\Delta \varepsilon$ changes when $\psi$ is chosen randomly.  Namely, the $\Delta \varepsilon$ can be reduced for instances of strong spin-orbit coupling, but not expanded.  Thus, we use a single value for the initial spin longitude ($\psi_o = 23.76^\circ$) in all of our simulations.

From these initial parameters, we evolve the secular time-dependant Hamiltonian that includes the canonical variable $\chi$ ($=\cos \varepsilon$) and spin longitude $\psi$ in the following equations of motion \citep{deSurgy1997,Saillenfest2019}:

\begin{align} \label{eqn:eom}
    \frac{\delta \psi}{\delta t} &= \frac{\gamma \chi}{\left(1-e^2\right)^{3/2}} - \frac{\chi}{\sqrt{1-\chi^2}}\left[ \mathcal{A}(t)\sin \psi + \mathcal{B}(t) \cos \psi \right] - 2\mathcal{C}(t)  \\ 
    \frac{\delta \chi}{\delta t} &= \sqrt{1-\chi^2}\left[\mathcal{B}(t)\sin\psi - \mathcal{A}(t)\cos \psi \right], 
\end{align}
\noindent where the functions $\mathcal{A}(t)$, $\mathcal{B}(t)$, and $\mathcal{C}(t)$ depend on the orbital evolution of the planet through $p = \sin(i/2)\sin \Omega$ and $q = \sin(i/2)\cos \Omega$ in the following relations\footnote{Note that \cite{Quarles2019} contains a typographical error in $\mathcal{B}(t)$, which has been corrected here.}:

\begin{align}
    \mathcal{A}(t) &= 2\left(\dot{q}+p\left(q\dot{p} - p\dot{q} \right) \right)/\sqrt{1-p^2-q^2},  \\
    \mathcal{B}(t) &= 2\left(\dot{p}-q\left(q\dot{p} - p\dot{q} \right) \right)/\sqrt{1-p^2-q^2}, \\
    \mathcal{C}(t) &= \left(q\dot{p} - p\dot{q} \right),
\end{align}

 \noindent using the numerical integration routines from the \texttt{scipy} library \citep{2020SciPy} within \texttt{python} in decade steps between each state that is recorded from a given n-body simulation.  In cases where the secular timescale is shorter than 10,000 yr, the steps are shortened to five years to ensure that changes in obliquity between output steps remain small ($\lesssim0.25^\circ$).

\subsection{Climate Model}
The most robust method for climate modeling is using a General Circulation Model (GCM), but such a method is computationally expensive \citep{Way2017} and consequently is not amenable to covering large swathes of parameter space.  Therefore, we employ a one-dimensional (1D) energy balance model (EBM) that incorporates distinctions between the heat capacities of land masses and oceans.  Specifically, we use a modified version of \texttt{VPLanet}{\footnote{Our modifications to the \texttt{POISE} module are included in the current version of the \texttt{VPLanet} repository (\url{https://github.com/VirtualPlanetaryLaboratory/vplanet}).}} \citep{Barnes2019} that directs the \texttt{POISE} module \citep{Deitrick2018b} to read from a file for the orbital and obliquity evolution described in Sections \ref{sec:orbit} and \ref{sec:obliq}.  To ensure the accuracy within our climate models, the orbital and obliquity evolution is sampled at either 5 or 10 year intervals.  The shorter 5 year interval is used for special cases when the secular evolution timescale is sufficiently fast ($<10,000$ yr), whereas the longer 10-year interval is used in all other cases.  The \texttt{POISE} module makes some simplifying assumptions where the planet rotation timescale is much shorter than the orbital period.  Those assumptions are valid for a spin precession constant $\gamma \gtrsim 1$\arcsec/yr because the orbital period of Earth-like planets in either HZ of $\alpha$ Centauri AB is more than 220 days.

The module \texttt{POISE} implements a one-dimensional EBM \citep{Budyko1969,Sellers1969} based on \cite{North1979} with several modifications specifically including ice {coverage} evolution (growth, melting, and flow).  Additionally, the module divides each latitudinal cell into land and ocean portions to produce a coupled set of equations to model the heat flow.  The equations for land and water depend on the parameterized latitude $x = \sin \phi$, where $\phi$ is the latitude, and are as follows:

\begin{align}
    C_L \frac{\partial T_L}{\partial t} - D\frac{\partial}{\partial x}(1-x^2)\frac{\partial T_L}{\partial x} + \frac{\nu}{f_L}(T_L - T_W) + I(x,T_L,t) = S(x,t)(1-\alpha(x,T_L,t)), \\
    m_d C_W \frac{\partial T_W}{\partial t} - D\frac{\partial}{\partial x}(1-x^2)\frac{\partial T_W}{\partial x} + \frac{\nu}{f_W}(T_W - T_L) + I(x,T_W,t) = S(x,t)(1-\alpha(x,T_W,t)), \label{eqn:EBM_water}
\end{align}
where the latitudinal temperature $T$ and heat capacity $C$ have subscripts corresponding to the land (\textit{L}) or water (\textit{W}) portion.  Equation \ref{eqn:EBM_water} contains an adjustable parameter $m_d$ that corresponds to the mixing depth within the oceans.   These equations describe a balance of energy flow ($C [\partial T/\partial t$]) with the outgoing longwave radiation $I(x,T,t)$, the incident insolation $S(x,t)$, and the planetary albedo $\alpha$.  There is an additional term that contains a parameter $\nu$, the latitudinal temperatures ($T_L$ and $T_W$), and the fraction $f$ due to land or water.  This term is used as a boundary condition within a latitudinal cell and the parameter $\nu$ adjusts the land–ocean heat transfer to reasonable values.  The total number of latitudinal cells $n_{\rm lat}$ is 151 ranging from --83.4$^\circ$S to 83.4$^\circ$N, where the equal steps are taken in the parameterized latitude $x$.  Note that latitude cells of
size $dx$ do not have equal width in latitude, but are equal in area \citep{Barnes2019}.

The incident insolation $S(x,t)$ relates to the latitude through the parameter $x = \sin \phi$ and time (or true longitude $\lambda$) through the declination $\delta = \psi \sin \lambda$ of the host star.  The true longitude marks the location of the planet within its orbit and can be easily calculated \citep{Brouwer1961,Berger1978} as long as changes to the semimajor axis and eccentricity are small over a single orbit.  The magnitude of the insolation $S_\star$ depends on the distance $r$ between the planet and its host star at each point within its orbit, where it is maximal and minimal during the planet's pericenter and apocenter passage, respectively.  The general form of $S(x,t)$ is

\begin{align}
    S(x,t) = \frac{L_\star}{4\pi^2 r^2}\left(H_o x \sin \delta + (1-x^2)\cos \delta \sin H_o \right),
\end{align}
which includes the stellar luminosity $L_\star$ and the hour angle $H_o$.  The hour angle calculation determines the day length for a given latitude, where many studies have provided derivations \citep{Laskar1993a,Armstrong2014,Quarles2020a} and a detailed discussion is given in the documentation for \texttt{VPLanet} \citep{Barnes2019}.

The planetary albedo is separated by surface type (land or water), temperature, and zenith angle.  Once the temperature drops below a critical value, then ice forms over the land or water and increases its albedo.  As a result the planetary albedo {(see Table \ref{tab:EBM_params})} is explicitly defined for land grid cells as

\begin{align}
    \alpha = \begin{cases}
    \alpha_L + 0.08{\rm P_2}(\sin Z), &{\rm if\; M_{ice} = 0\; and\; T_L>-2^\circ C} \\
    \alpha_{ice}, &{\rm if\; M_{ice} > 0\; or\; T_L\leq-2^\circ C},
    \end{cases}
\end{align}

\noindent while for water grid cells it is

\begin{align}
    \alpha = \begin{cases}
    \alpha_W + 0.08{\rm P_2}(\sin Z), &{\rm if\; T_W>-2^\circ C} \\
    \alpha_{ice}, &{\rm if\; T_L\leq-2^\circ C}
    \end{cases}
\end{align}
{\cite{Shields2013} showed how the host star's spectral type can affect potential climates through the ice-albedo feedback for FGKM stars, where Earth-analogs orbiting G or K dwarfs exhibit similar feedbacks.  However, \cite{Wilhelm2021} showed that the orbital period within the HZ is a bigger effect, which confirms Milankovitch's
hypothesis that inter-annual ice coverage depends most strongly on the strength and duration of summer melting.} These simulations assume an Earth-analog planet {in the HZ of either $\alpha$ Cen A or $\alpha$ Cen B}, where specifically the atmospheric chemical composition {(N$_2$/H$_2$O/CO$_2$)} and the land/water fraction {(25\% land and 75\% water)} is nearly identical to Earth values.  The land/water fraction is constant across latitudes, where the effect of geography is beyond the scope of this work.  Thus, we use the linear parameterized outgoing longwave radiation function
\begin{equation} 
    I = A + BT 
\end{equation}
that is implemented in \texttt{VPLanet}, and depends upon the surface temperature $T$, where  values for Earth are adopted from \cite{North1979}: $A = 203.2$ W m$^{-2}$ and $B = 2.09$ W $m^{-2}$ $^\circ$C$^{-1}$.  

We are primarily interested in the extent of ice {coverage} through Milankovitch cycles because of the interplay between ice-albedo feedback with obliquity variations.  \texttt{VPLanet} models the ice accumulation and ablation in a fashion similar to \cite{Armstrong2014} using the surface temperature $T$ and freezing temperature $T_{\rm freeze}$, according to the formula in \cite{Deitrick2018b}:
\begin{equation} \label{eqn:ice_growth}
    \frac{dM_{\rm ice}}{dt} = \frac{2.3\sigma}{L_h}\left[T_{\rm freeze}^4 - (T + T_{\rm freeze})^4\right],
\end{equation}
where $M_{\rm ice}$ is the surface mass density of ice, $\sigma = 5.67 \times
10^{-8}$ W m$^{-2}$ K$^{-4}$ is the Stefan-Boltzmann constant, $L_h$ is the
latent heat of fusion of ice, $3.34 \times 10^5$ J kg$^{-1}$, and $T_{\rm freeze} = 273.15$ K. The factor of 2.3 in Equation \ref{eqn:ice_growth} is used to scale the melt rate to roughly Earth values of 3 mm $^\circ$C$^{-1}$ day$^{-1}$.  Once the land temperature is below freezing, then ice accumulates on land at a constant rate $r_{\rm snow}$.  \texttt{VPLanet} allows for modeling of ice flows that modify the height of the ice and bedrock depression, but these effects are beyond the scope of this work.  The input parameters used for \texttt{VPLanet} are summarized in Table \ref{tab:EBM_params}.  {All of our climate simulations begin with an ice-free planet.}

% \begin{deluxetable}{lcc}
% \tablecolumns{3}
% \tablecaption{Parameters for the EBM and ice sheet model in \texttt{VPLanet}. \label{tab:EBM_params}}
% \tablehead{\colhead{Variable} & \colhead{Value} & \colhead{Units}}
% \startdata
% C$_L$ &  $1.55\times 10^7$ & J m$^{-2}$ K$^{-1}$ \\
% C$_W$ &  $4.428\times 10^6$ & J m$^{-3}$ K$^{-1}$ \\
% $m_d$ & 70 & m \\
% $D$ & 0.58 & W m$^{-2}$ K$^{-1}$ \\
% $\nu$ & 0.8 & -- \\
% $A$ & 203.3 & W m$^{-2}$ \\
% $B$ & 2.09 & W m$^{-2}$ K$^{-1}$ \\
% $\alpha_{\rm L}$ & 0.363 & -- \\
% $\alpha_{W}$ & 0.263 & -- \\
% $\alpha_{\rm ice}$ & 0.6 & -- \\
% $f_L$ & 0.34 & -- \\
% $f_W$ & 0.66 & -- \\
% $T_{\rm freeze}$ & 273.15 & K \\
% $L_h$ & $3.34 \times 10^5$ & J kg$^{-1}$ \\
% $r_{\rm snow}$ & $2.25\times10^{-5}$ & kg m$^{-2}$ s$^{-1}$ \\
% $n_{\rm lat}$ & 151 & --
% \enddata
% \end{deluxetable}
\begin{table}
    \caption{Parameters for the EBM and ice model in \texttt{VPLanet}. \label{tab:EBM_params}}
    \centering
    \begin{tabular}{ccc}
        \hline
        Variable & Value & Units \\
        \hline
         C$_L$ &  $1.55\times 10^7$ & J m$^{-2}$ K$^{-1}$ \\
        C$_W$ &  $4.428\times 10^6$ & J m$^{-3}$ K$^{-1}$ \\
        $m_d$ & 70 & m \\
        $D$ & 0.58 & W m$^{-2}$ K$^{-1}$ \\
        $\nu$ & 0.8 & -- \\
        $A$ & 203.3 & W m$^{-2}$ \\
        $B$ & 2.09 & W m$^{-2}$ K$^{-1}$ \\
        $\alpha_{\rm L}$ & 0.363 & -- \\
        $\alpha_{W}$ & 0.263 & -- \\
        $\alpha_{\rm ice}$ & 0.6 & -- \\
        $f_L$ & 0.34 & -- \\
        $f_W$ & 0.66 & -- \\
        $T_{\rm freeze}$ & 273.15 & K \\
        $L_h$ & $3.34 \times 10^5$ & J kg$^{-1}$ \\
        $r_{\rm snow}$ & $2.25\times10^{-5}$ & kg m$^{-2}$ s$^{-1}$ \\
        $n_{\rm lat}$ & 151 & -- \\
        \hline
    \end{tabular}
\end{table}

\section{An Earth-analog orbiting \texorpdfstring{$\alpha$}{alpha} Centauri A} \label{sec:aCenA}
The orbital and spin evolution of an Earth-analog orbiting either star in $\alpha$ Cen AB differs due to the planetary semimajor axis corresponding to the inner edge of the host star's HZ and the mass of the planet-hosting star.  In this section, the investigation focuses on a planet orbiting $\alpha$ Cen A inclined by 10$^\circ$ (relative to the binary orbital plane) at its forced eccentricity and holds the initial planetary orbit constant between simulations, while exploring a range of configurations for the initial spin state with respect to the precession constant $\gamma$ and initial obliquity $\varepsilon_o$.  Section \ref{sec:aCenB} explores a similar parameter space, but expands the our study to include three values for the planetary inclination relative to the binary orbital plane.

\subsection{Factors Affecting Milankovitch Cycles} \label{sec:aCenA_temp}
\begin{figure}
    \centering
    \includegraphics[width=\linewidth]{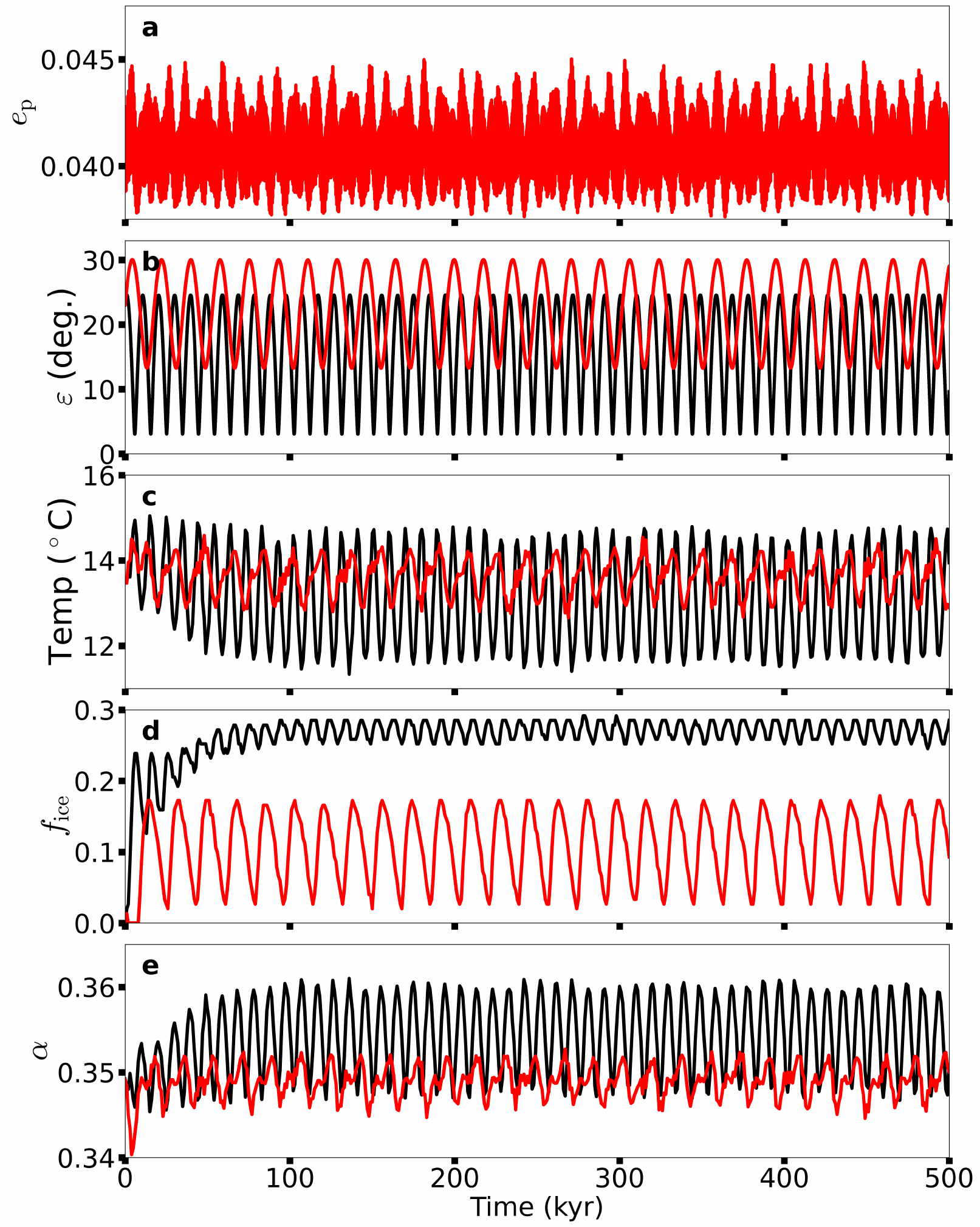}
    \vspace{-3ex}
    \caption{Evolution over a 500 kyr timescale of the (a) planetary eccentricity $e_{\rm p}$, (b) obliquity $\varepsilon$, (c) global surface temperature, (d) global ice fraction $f_{\rm ice}$, and (e) global albedo $\alpha$.  The planetary orbit begins near its forced eccentricity at the inner edge of $\alpha$ Cen A's HZ with an inclination of 10$^\circ$.  The initial planetary obliquity $\varepsilon$ is similar to modern Earth obliquity (23$^\circ$), where the precession constant $\gamma$ is either 10\arcsec/yr (black) or 85\arcsec/yr (red).  The former precession constant represents a modern, moonless Earth ($P_{\rm rot}\sim24$ hr) and the latter represents a more rapidly spinning Earth ($P_{\rm rot}\sim14$ hr).  The planetary orbit between these simulations is identical and thus, the evolution for the eccentricity overlaps exactly. \label{fig:EBM_ts_A} }
\end{figure}

The planetary orbit in our simulations begins at the forced eccentricity ($e_F\approx 0.04$), which warms and cools the whole planet over an orbit ($\sim$1.2 yr), however this short period variation largely averages out over timescales relevant for global climate.  Moreover, starting at the forced eccentricity minimizes variations in the osculating eccentricity \citep{Quarles2018a}.  Figure \ref{fig:EBM_ts_A} illustrates the evolution in the planetary eccentricity $e$, obliquity $\varepsilon$, global temperature, global ice fraction $f_{\rm ice}$, and the global albedo $\alpha$ for two simulations with Earth-like parameters but with different rotation periods ($\sim$24 hr (black) and $\sim$14 hr (red)).  Figure \ref{fig:EBM_ts_A}a shows a maximum eccentricity variation $\Delta e\approx 0.008$ over 500 kyr and applies to all of our simulations in Section \ref{sec:aCenA}.

A planet representing a modern, moonless Earth has a precession constant of 10\arcsec/yr (black) in Fig. \ref{fig:EBM_ts_A}, which results in quite rapid variations in the planetary obliquity $\Delta\varepsilon \sim 22^\circ$ (Fig. \ref{fig:EBM_ts_A}b).  If the planet has a precession constant of 85\arcsec/yr (red; Fig. \ref{fig:EBM_ts_A}b), the obliquity variation is reduced to $\Delta\varepsilon \sim 18^\circ$.  This reduction is a result of a stronger spin-orbit coupling where the spin precession more closely matches the orbital precession as described in \cite{Quarles2019}.  The effect of the spin-orbit coupling manifests in the global surface temperature (Fig. \ref{fig:EBM_ts_A}c) through the frequency of the temperature variations, where the rapid rotator (red) has a lower frequency than the slower rotator (black).  Note that the global surface temperature is calculated as a weighted average over the latitude.  In addition to a difference in frequency, the slower rotator (black) undergoes larger variations ($\sim3.5^\circ$C) in the global mean surface temperature compared to the faster rotator ($\sim1.9^\circ$C).  The larger oscillation in surface temperature is caused (in part) by the more extreme obliquity variation, where the growth of ice {coverage}, the excursion to low obliquity ($\varepsilon <10^\circ$), and changes in albedo all contribute to a higher variation in the global surface temperature. \texttt{POISE} is initialized to a steady state temperature, but Fig. \ref{fig:EBM_ts_A}c shows that it is only a local equilibrium because of the ice growth (Figs. \ref{fig:EBM_ts_A}d and \ref{fig:EBM_ts_A}e).  In general, the growth of ice {coverage} increase the surface albedo, which causes a decrease in the surface temperature.  Such a trend is shown in Figs. \ref{fig:EBM_ts_A}c--\ref{fig:EBM_ts_A}e for the first $\sim$100 kyr until a steady state is achieved in the global ice fraction.  Consequently, we ignore the first 100 kyr when computing the variations $\Delta$ in our analysis of the full range of initial planetary spin states (i.e., varying $\gamma$ \& $\varepsilon_o$).

\begin{figure}
    \centering
    \includegraphics[width=\linewidth]{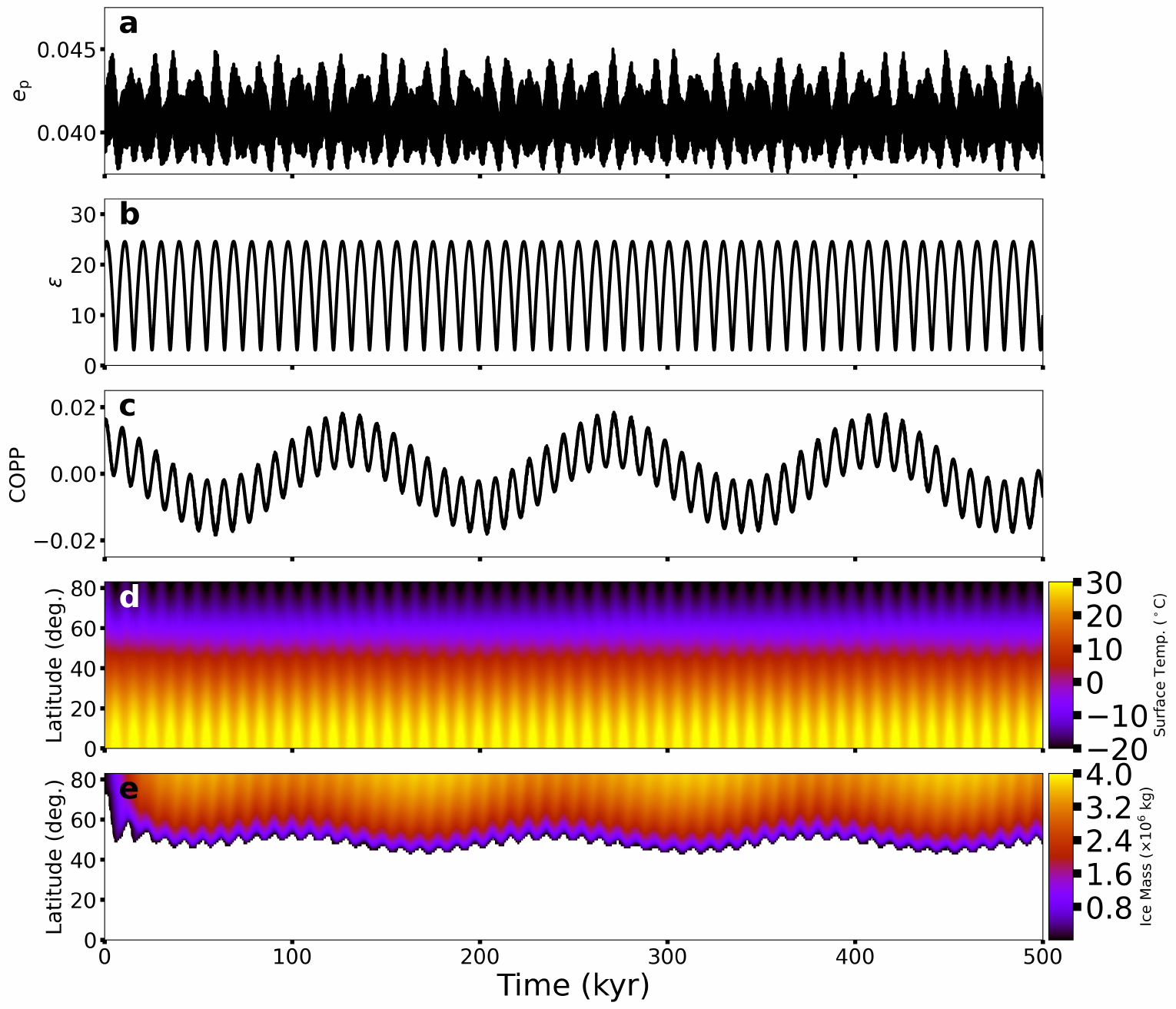}
    \caption{Simulation for 500 kyr of an Earth-analog ($\varepsilon=23^\circ$) orbiting $\alpha$ Cen A with a spin precession constant equal to 10\arcsec/yr showing the evolution in the (a) planetary eccentricity $e_{\rm p}$, (b) obliquity $\varepsilon$, (c) {Climate Obliquity Precession Parameter (COPP, $e_{\rm p}\sin{(\varepsilon)}\sin(\varpi_{\rm p} + \psi)$)}, (d) latitudinal surface temperature, and (e) latitudinal ice mass.  The white cells in (e) represent latitudes with very little ice accumulation ($<10^4$ kg) or ``ice free''. \label{fig:EBM_ts_A_10} }
\end{figure}

Previous works \citep[e.g.,][]{Armstrong2014,Deitrick2018b} used a parameter that combines the effect of the planetary eccentricity, obliquity, and their associated angles (longitude of pericenter $\varpi_{\rm p}$ and spin longitude $\psi$), which is dubbed the Climate {Obliquity} Precession Parameter {(COPP, $e_p \sin{(\varepsilon})\sin{(\varpi_{\rm p}+\psi)}$)}.  Figure \ref{fig:EBM_ts_A_10} illustrates the {COPP} (Fig. \ref{fig:EBM_ts_A_10}c) for the modern, moonless Earth-analog with a spin precession constant set to 10\arcsec/yr (black curves in Fig. \ref{fig:EBM_ts_A}) along with the surface temperature (Fig. \ref{fig:EBM_ts_A_10}d) and ice mass (Fig. \ref{fig:EBM_ts_A_10}e) relative to a given latitude in the northern hemisphere.  The {COPP} is bounded by the planet's eccentricity {and obliquity, while retaining} an imprint of {both variations}, but there is a longer ($\sim$140 kyr) cycle from the combination of orbital and spin precession.  Milankovitch cycles for the Earth arise from the combination of the short and long timescales from the spin and orbital precession.  Similar trends occur within our numerical simulation for a planet orbiting $\alpha$ Cen A as shown by the light/dark variations for the surface temperature and ice mass in Figure \ref{fig:EBM_ts_A_10}d and \ref{fig:EBM_ts_A_10}e, respectively.  The white cells in Fig. \ref{fig:EBM_ts_A_10}e mark latitudes with very little ice mass ($<10^4$ kg) that we categorize as ``ice free'' because the transition from no ice to significant ice {coverage} is stark, partly due to our limited resolution in latitude.

\begin{figure}
    \centering
    \includegraphics[width=\linewidth]{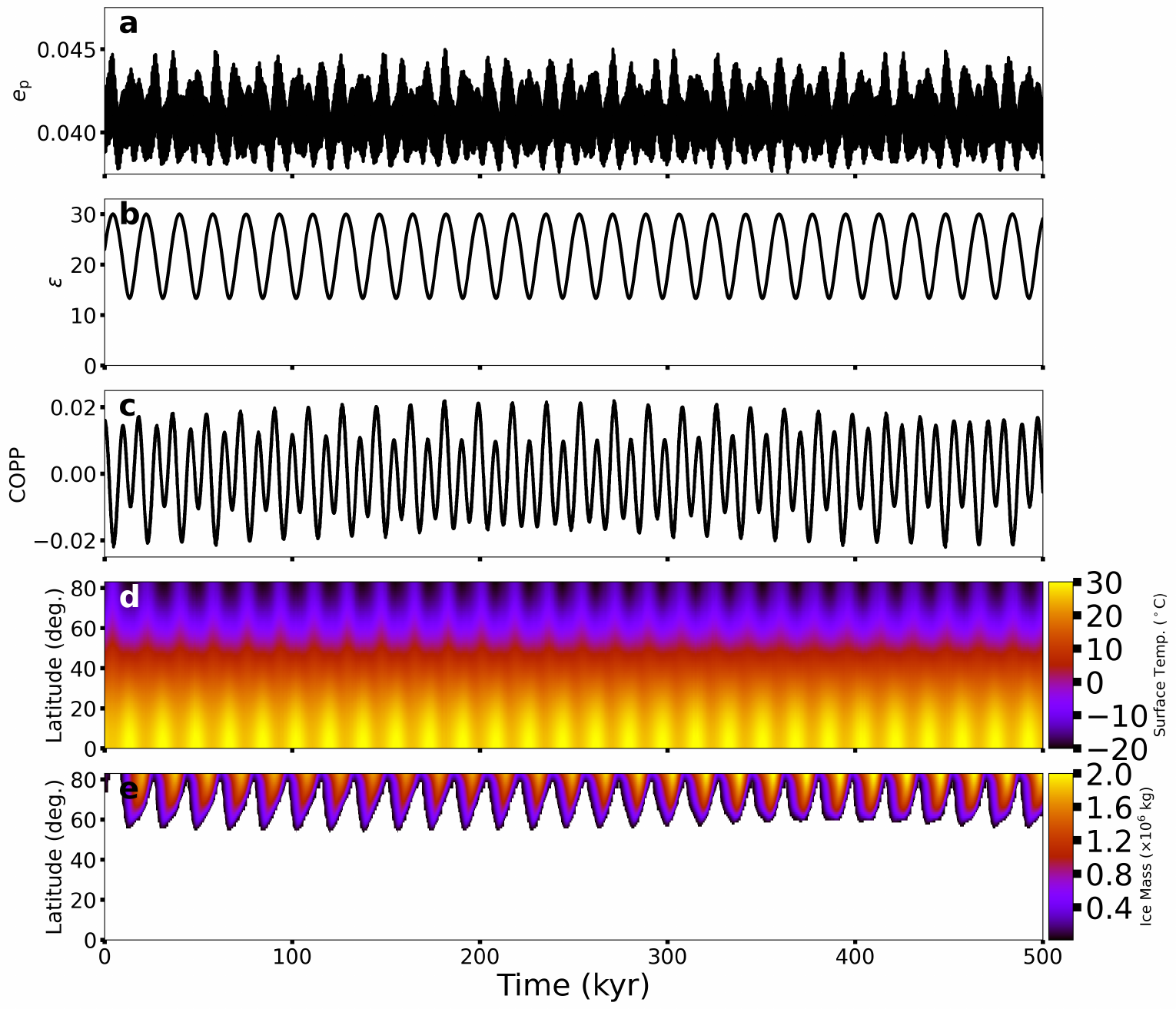}
    \caption{Similar to Fig. \ref{fig:EBM_ts_A_10}, but for an Earth-analog with a spin precession constant equal to 85\arcsec/yr. \label{fig:EBM_ts_A_85} }
\end{figure}

To disentangle the contribution of orbital or spin precession, we produce a similar analysis, but for a larger spin precession constant (85\arcsec/yr) in Figure \ref{fig:EBM_ts_A_85}.  The orbital evolution between the two cases (Figs. \ref{fig:EBM_ts_A_10} and \ref{fig:EBM_ts_A_85}) is identical and thus the changes to the {COPP} (Fig. \ref{fig:EBM_ts_A_85}c) are largely due to the higher spin precession.  The Milankovitch cycles are more dramatic, where the higher latitudes undergo enough warming for the ice {coverage} to periodically retreat by $\sim$15$^\circ$ poleward in latitude.

\subsection{Obliquity and Global Surface Temperature Variations} \label{sec:param_A}
\begin{figure}
    \centering
    \includegraphics[width=0.6\linewidth]{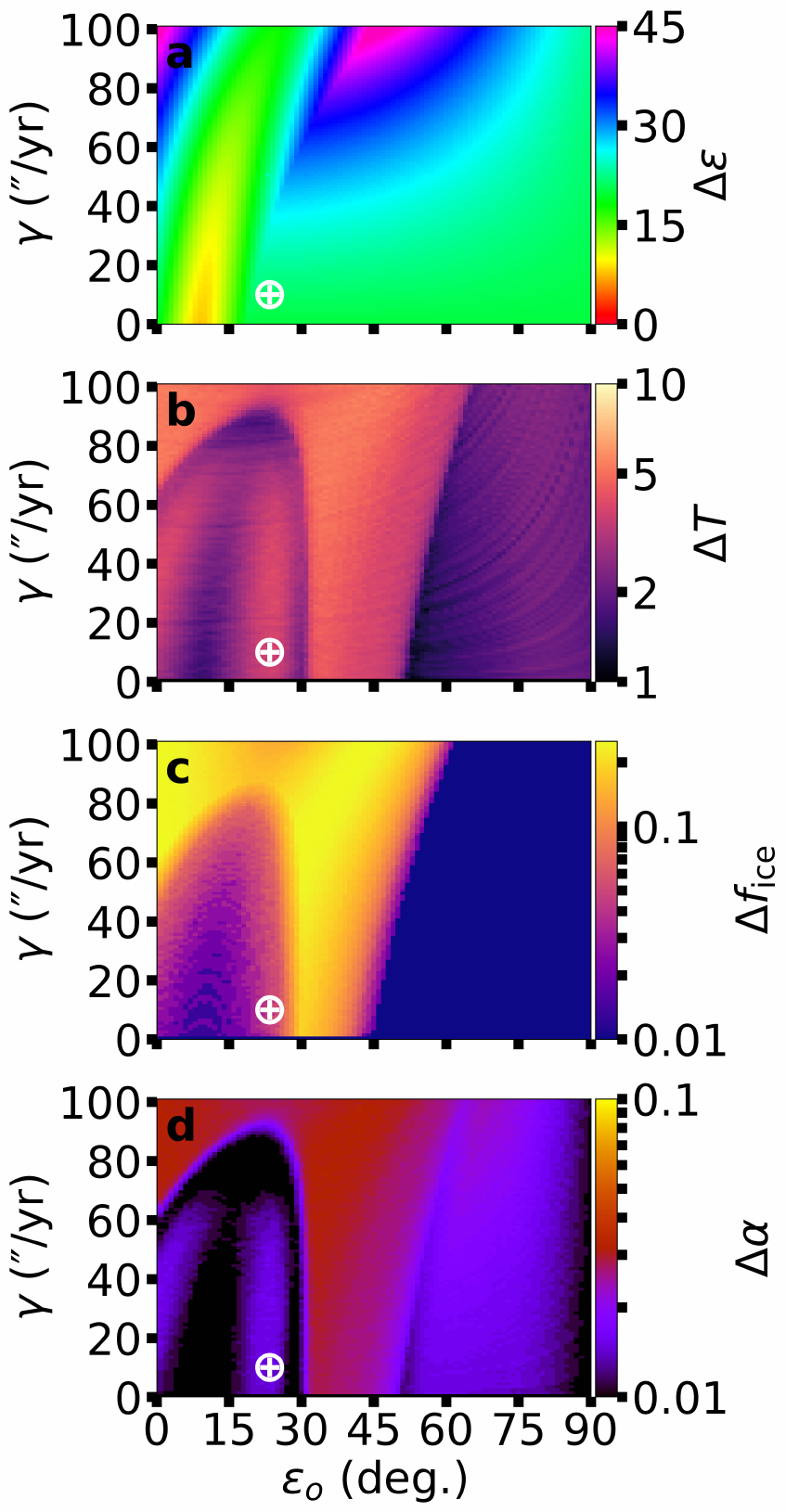}
    \caption{The variation of the (a) obliquity $\Delta \varepsilon$, (b) global surface temperature $\Delta T$, (c) global ice fraction $\Delta f_{\rm ice}$, and (d) global albedo $\Delta \alpha$ using simulations with an EBM over 400 kyr for an Earth-analog orbiting $\alpha$ Cen A, where the planet's orbit is initially inclined by 10$^\circ$ relative to the binary orbit.  The initial spin is varied with respect to the spin precession constant $\gamma$ and the initial obliquity $\varepsilon_o$.  The white $\oplus$ symbol designates conditions ($\varepsilon_o=23^\circ$, $\gamma = 10$\arcsec/yr, $P_{\rm rot}\sim24$ hr) for a modern, moonless Earth.  \label{fig:aCenA_all} }
\end{figure}

For a broader view, we perform simulations considering a prograde rotating planet ($\varepsilon=0^\circ-90^\circ$) and a wide range of precession constants ($\gamma = 0$--100\arcsec/yr) in Figure \ref{fig:aCenA_all}, where the variations of each parameter are color-coded. {We note that the maximum obliquity does not exceed $90^\circ$ in any of our simulations and the obliquity range when $\varepsilon_o \sim 90^\circ$ extends to lower obliquity.} The white $\oplus$ symbol denotes initial conditions for a modern, moonless Earth-analog ($\gamma = 10$\arcsec/yr and $\varepsilon_o = 23^\circ$).  For all of these simulations, the planet begins at the inner edge of $\alpha$ Cen A's HZ so that the planet receives an Earth-equivalent amount of radiative flux at the top of its atmosphere and the planetary orbit begins inclined 10$^\circ$ relative to the binary orbital plane, where we expect an obliquity variation $\Delta \varepsilon\approx15-20^\circ$ from the induced nodal precession on the planetary orbit.  This explains much of the parameter space (light green regions) in Figure \ref{fig:aCenA_all}a.  

Stronger spin-orbit coupling produces a strip of lower than expected obliquity variation ({yellow}) similar to the case for $\alpha$ Cen B \citep{Quarles2019} {, corresponding to the region close to the Cassini state}.  The wide separation of secular modes within $\alpha$ Cen A's HZ prevent the large obliquity variations seen for fast rotators ($\gamma \sim 80$--100\arcsec/yr) in $\alpha$ Cen B's HZ examined in \cite{Quarles2019}.  The relatively mild obliquity variation for an Earth-analog orbiting $\alpha$ Cen A correlates with a mild variation ($\Delta T$) in the global surface temperature with a maximum of 5.6$^\circ$C in Fig. \ref{fig:aCenA_all}b.  The strip of lower obliquity variation also induces some of the smallest surface temperature variations for low initial obliquity.  Intermediate values of initial obliquity (30$^\circ$--50$^\circ$) exhibit the maximum surface temperature variation, while the highest obliquities have much lower $\Delta T$.

\subsection{Ice Fraction and Surface Distribution} \label{sec:aCenA_ice}
Two strongly correlated factors that affect a planet's global surface temperature, and thereby its habitability, are the global ice fraction ($\Delta f_{\rm ice}$) and albedo ($\Delta \alpha$) variations.  Our simulations begin completely ice free, where a dynamic equilibrium in ice coverage is established within $\sim$100 kyr.  Our calculations of the variation exclude the first 100 kyr to remove this bias and so we can measure the overall variation relative a steady-state.  Figure \ref{fig:aCenA_all}c and \ref{fig:aCenA_all}d show their highest variations are correlated with the global surface temperature (Fig. \ref{fig:aCenA_all}b).  The largest variation in global ice fraction is $\sim$0.2--0.3, resulting in a variation in the global albedo of $\sim0.03$.  There is also a similar trend with respect to the initial obliquity as seen in the surface temperature variation.  The global variation provides a broad overview, but is insufficient in the necessary details for Milankovitch cycles.

From our simulations, we categorize the ice distribution into 4 states: 1) ice free, 2) ice caps, 3) ice belt, and 4) snowball, where each of these are determined after the removal of the first 100 kyr.  The ice free category describes planets where ice does not accumulate year-to-year, although seasonal ice is still possible.  It is possible that ice accumulates and ablates between the 1000 yr interval for our simulation outputs, but we consider this scenario fine-tuned and unlikely.  The snowball category designates when ice {coverage} grows to cover the entire planet, which persists throughout the simulation.  There are additional options available in \texttt{POISE} (e.g., more sophisticated heat diffusion or CO$_2$ partial pressures) that could allow for snowballs to thaw, but this is beyond the scope of this work.  Ice caps or ice belt are states with ice that extend from the pole down to 30$^\circ$ N or from the equator up to 30$^\circ$ N \citep{Williams2003,Rose2017}, respectively.  Since the planetary eccentricity in our simulations remains nearly circular, the ice distributions are largely symmetric relative to the equator.  \cite{Armstrong2014} showed that this is not the case for highly inclined and eccentric planets, where more sophisticated categorization criteria than what we employ would be required in those circumstances (e.g., orbits closer to the binary mean motion resonances).  Our model does permit transitions between the ice free, ice caps, and ice belt states, which is the essence of Earth-like Milankovitch cycles.  {The Earth science community will refer to an \emph{ice cap} by the extent of ice coverage on land (i.e., less than 50,000 km$^2$).  Our usage of ice cap is different, where ice caps are states with ice that extends from the pole to at least 30$^\circ$, without a delineation between land/ocean coverage.}  

\begin{figure}
    \centering
    \includegraphics[width=0.6\linewidth]{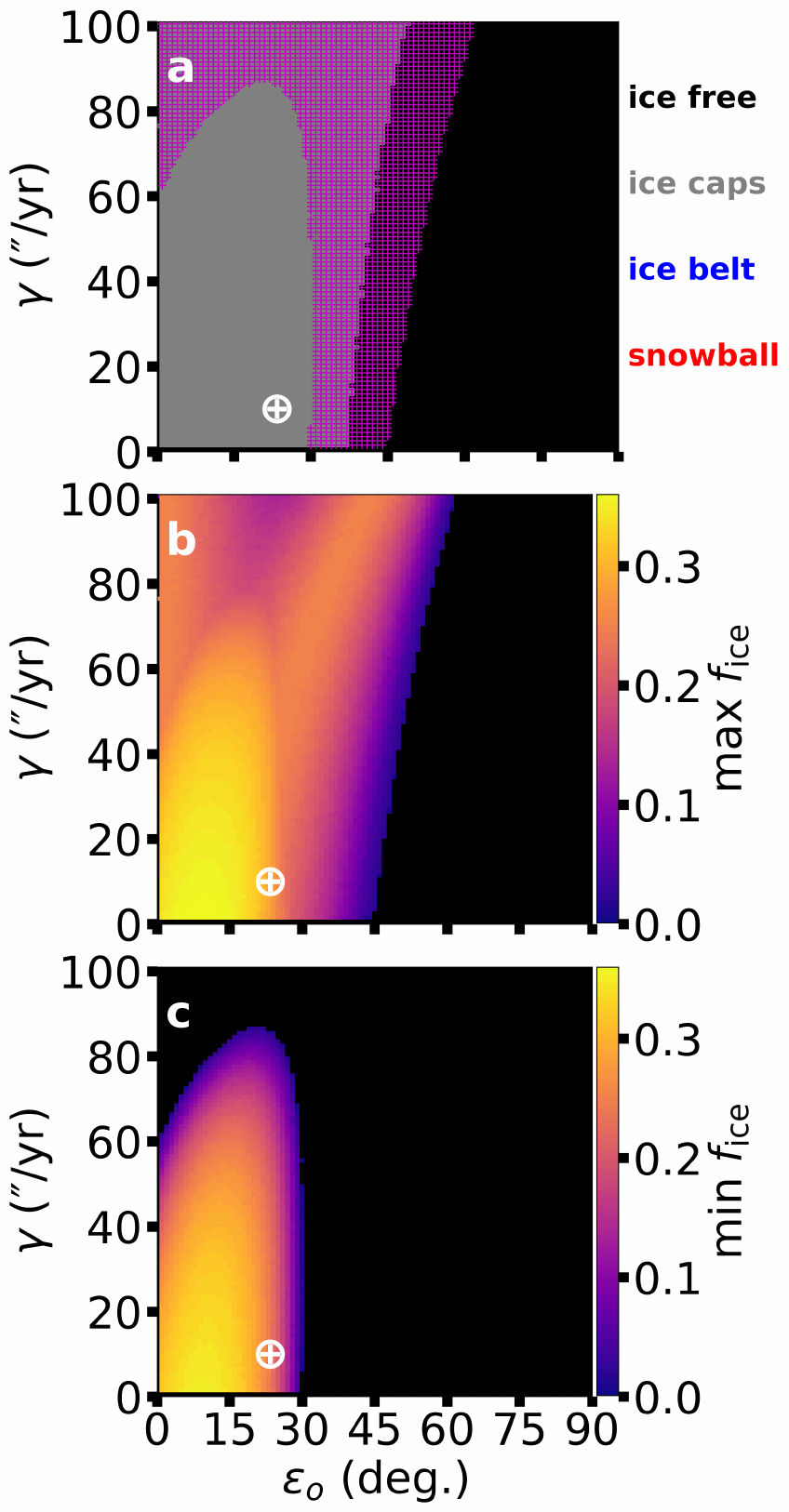}
    \caption{Categorization of the most frequent ice distribution state (ice free, ice caps, ice belt, or snowball) within each simulation in panel (a).  Panels (b) and (c) illustrate the maximum and minimum global ice fraction, respectively.  The {magenta hatched region} plotted in (a) {represents the cases where} the planet oscillates between two ice distribution states.  {Labels for ice belt and snowball are provided for consistency with Figure \ref{fig:aCenB_ice}, although none of the simulations represented here enter either of these states.}  \label{fig:aCenA_ice} }
\end{figure}

Figure \ref{fig:aCenA_ice} illustrates our results in Fig. \ref{fig:aCenA_ice}a using the ice distribution categorization scheme in addition to the maximum (Fig. \ref{fig:aCenA_ice}b) and minimum (Fig. \ref{fig:aCenA_ice}c) global ice fraction.  The ice distributions in Fig. \ref{fig:aCenA_ice}a are color-coded, where the {magenta hatched region denotes} simulations where the planet oscillates between an ice free state and one with polar ice caps.  {The ice belt and snowball categories are present for consistency with a similar figure (Figure \ref{fig:aCenB_ice}), although these categories were not observed for any of the simulations for $\alpha$ Cen A}.  The white $\oplus$ symbol denotes initial conditions representing a modern, moonless Earth as previously indicated in Fig. \ref{fig:aCenA_all}.  Figure \ref{fig:aCenA_ice}a shows two states (ice free or ice caps), where permanent ice caps (solid gray) are possible for low obliquities and a broad range in spin precession constant.  Surrounding this region ({magenta hatches}), ice caps persist most of the time and interrupted by relatively brief ice free periods.  This can also be deduced (to some degree) from spotting the differences between Figs. \ref{fig:aCenA_ice}b and \ref{fig:aCenA_ice}c, which indicate the respective maximum and minimum global ice fraction attained over a simulation or using the global ice fraction variation in Fig. \ref{fig:aCenA_all}c.  Alternatively, there are scenarios when the polar caps state is transient and an ice free state is more common, which is indicated by the {magenta hatching} on the black background in Fig. \ref{fig:aCenA_ice}a.  In this regime, the ice fraction strongly depends on the initial obliquity and more weakly depends on the spin precession constant.  For $\epsilon_o>60^\circ$, each model produce an ice free planet, which is due to the weak planetary spin-orbit coupling that induces up to a 20$^\circ$ variation in obliquity on a $\sim$10,000 year timescale (i.e., secular orbital eccentricity variations).  One might expect an ice belt to form if a planet begins at high obliquity \citep{Williams2003,Rose2017,Kilic2018}, but the planet's obliquity evolves quickly (within $\sim$5,000 years) to lower values such that the ice ablation rate matches the accumulation and no significant growth in ice {coverage} can occur.

\section{An Earth-analog orbiting \texorpdfstring{$\alpha$}{alpha} Centauri B} \label{sec:aCenB}
An Earth-analog orbiting $\alpha$ Cen B can experience similar outcomes to those discussed in the Section \ref{sec:aCenA} due to the approximate mass symmetry of the host binary.  However, $\alpha$ Cen B is less luminous than $\alpha$ Cen A, which places $\alpha$ Cen B's HZ at smaller separations, thereby altering the secular forcing frequency from the binary companion.  In addition, the planetary spin precession frequency for 24 hr rotator increases from $\sim$10\arcsec/yr to $\sim$46\arcsec/yr (see \cite{Quarles2019} for more details).  We expand our investigation with an Earth-analog orbiting $\alpha$ Cen B to include three planetary inclinations (2$^\circ$, 10$^\circ$, and 30$^\circ$) so that we sample a broader range of potentially habitable orbital configurations.

\subsection{Factors Affecting Milankovitch Cycles}
\begin{figure}
    \centering
    \includegraphics[width=\linewidth]{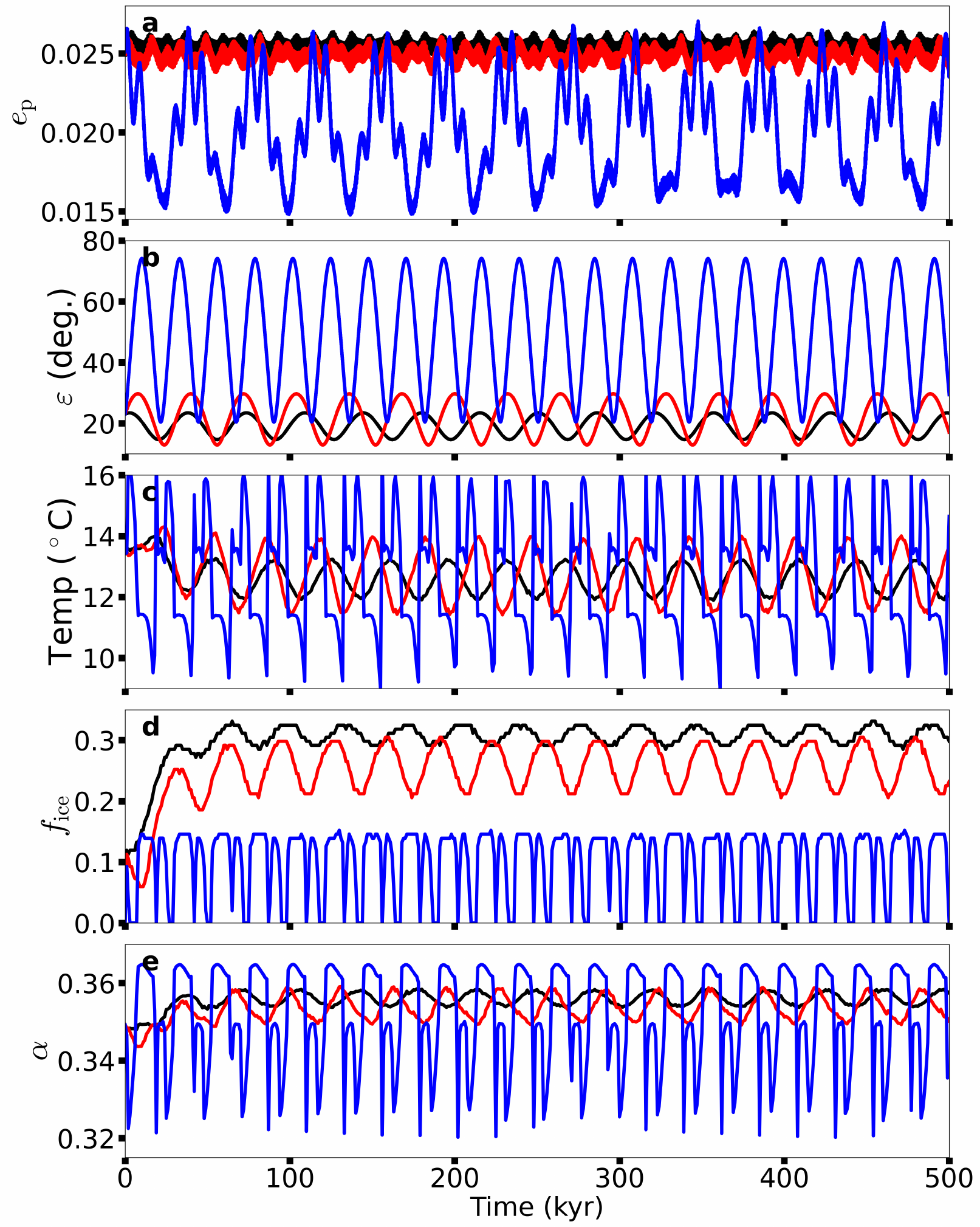}
    \caption{Evolution over a 500 kyr timescale of the (a) planetary eccentricity $e_{\rm p}$, (b) obliquity $\varepsilon$, (c) global surface temperature, (d) global ice fraction $f_{\rm ice}$, and (e) global albedo $\alpha$.  The planetary orbit begins near its forced eccentricity at the inner edge of $\alpha$ Cen B's HZ with an inclination of 2$^\circ$ (black), 10$^\circ$ (red), and 30$^\circ$ (blue).  The initial planetary obliquity $\varepsilon$ and precession constant $\gamma$ are similar in value to a modern, moonless Earth (23$^\circ$ and 46\arcsec/yr, respectively). \label{fig:EBM_ts}}
\end{figure}

The evolution of the planetary eccentricity and obliquity can correlate with changes in the global mean surface temperature through seasonal changes to the incoming radiation.  In addition, the global mean ice fraction $f_{\rm ice}$ and global mean albedo affect the global surface temperature by modifying the contribution of the outgoing radiation.  Figure \ref{fig:EBM_ts} demonstrates these variations for a modern, moonless Earth-analog ($\varepsilon_o=23^\circ$, $\gamma=46$\arcsec/yr) with an inclination $i_{\rm p}$ of 2$^\circ$ (black), 10$^\circ$ (red), and 30$^\circ$ (blue).  For 30$^\circ$ (blue; Fig. \ref{fig:EBM_ts}a), the eccentricity variation is larger, but the magnitude of the planetary eccentricity remains small.  The obliquity variation increases as a function of $i_{\rm p}$ in Fig.  \ref{fig:EBM_ts}b, where we expect the maximum variation to scale with twice the planetary inclination ($\Delta \varepsilon\sim 2i_{\rm p}$).  Similar to Fig. \ref{fig:EBM_ts_A}, the planet's eccentricity and obliquity variation induces oscillations in the global surface temperature (Fig. \ref{fig:EBM_ts}c), where the temperature changes are mild for low mutual inclination ($\lesssim10^\circ$; red and black).

A planetary orbit with a larger mutual inclination ($30^\circ$; blue) undergoes apsidal and nodal precession, which allows for stark changes in surface temperature and a larger variation.  Planets orbiting $\alpha$ Cen B reach a steady state with respect to the global ice fraction $f_{\rm ice}$ and albedo $\alpha$ after the first 100 kyr.  Interestingly, Fig. \ref{fig:EBM_ts}d shows that the low mutual inclination ($\lesssim10^\circ$; red and black) cases have a higher mean $f_{\rm ice}$ than the larger mutual inclination ($30^\circ$; blue).  This subsequently manifests in a higher average global albedo (Fig. \ref{fig:EBM_ts}e).  The larger mutual inclination ($30^\circ$; blue) planet has a maximum $f_{\rm ice}\sim 0.12$, but the global albedo can fluctuate over a 1000 yr interval between $\sim$0.32--0.36.  In fact, the larger mutual inclination ($30^\circ$; blue) planet experiences transient ice free states (Fig. \ref{fig:EBM_ts}d) that correlate with the mean value of the assumed albedo parameters for land and water (i.e., [$\alpha_{\rm L} + \alpha_{\rm W}$]/2; Table \ref{tab:EBM_params}).  The mean global albedo in the larger mutual inclination ($30^\circ$; blue) simulation is lower (Fig. \ref{fig:EBM_ts}e) in response to the smaller global ice fraction (Fig. \ref{fig:EBM_ts}d).

\begin{figure}
    \centering
    \includegraphics[width=\linewidth]{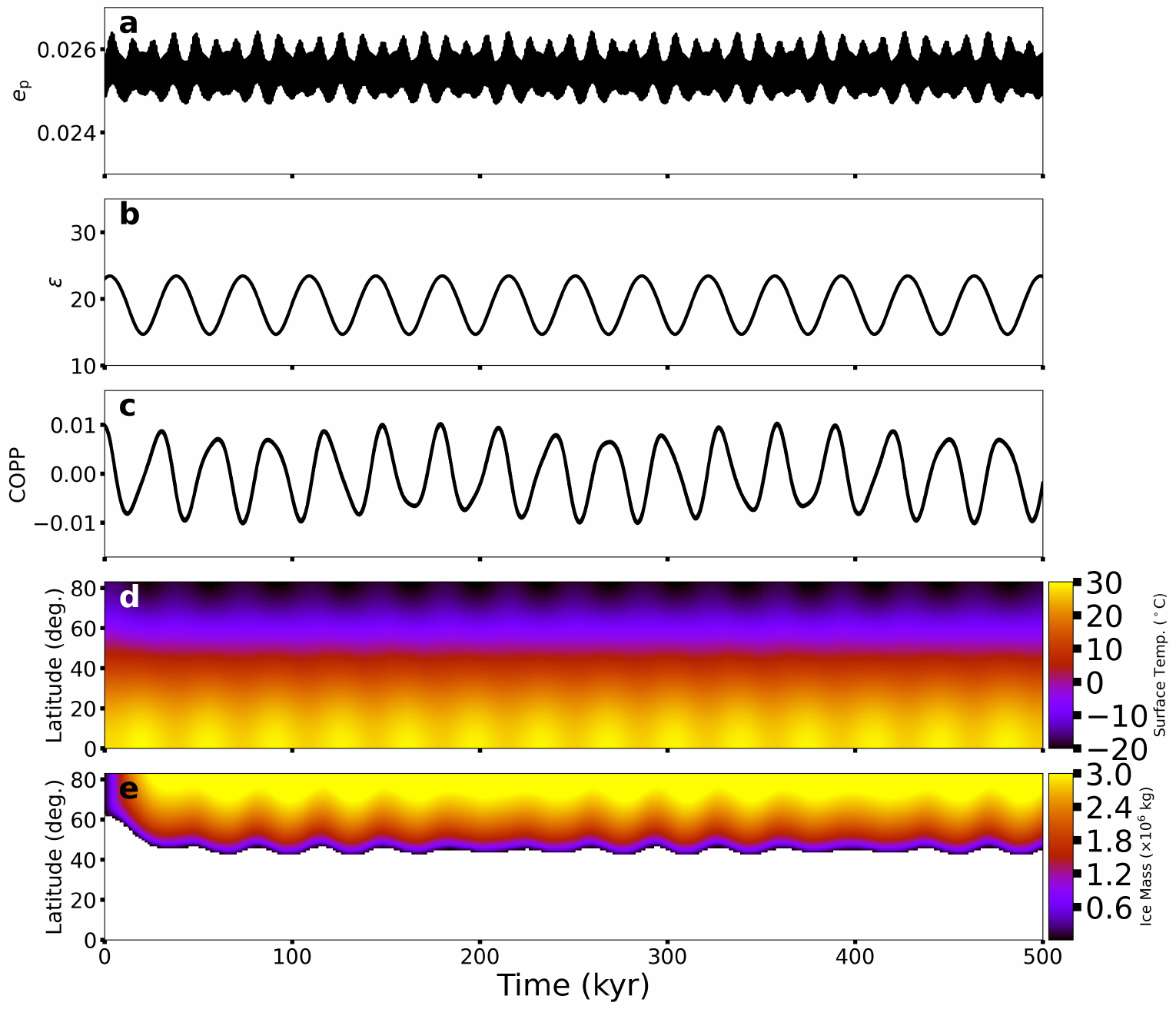}
    \caption{Simulation of an Earth-analog ($\varepsilon=23^\circ$) initially inclined 2$^\circ$ above the binary orbital plane showing the evolution in the (a) planetary eccentricity $e_{\rm p}$, (b) obliquity $\varepsilon$,  (c) {Climate Obliquity Precession Parameter (COPP; $e_{\rm p}\sin{(\varepsilon)}\sin(\varpi_{\rm p} + \psi)$)}, (d) latitudinal surface temperature, and (e) latitudinal ice mass.  The white cells in (d) represent latitudes with very little ice accumulation ($<10^4$ kg) or ``ice free''. \label{fig:EBM_ts_B_2} }
\end{figure}

\begin{figure}
    \centering
    \includegraphics[width=\linewidth]{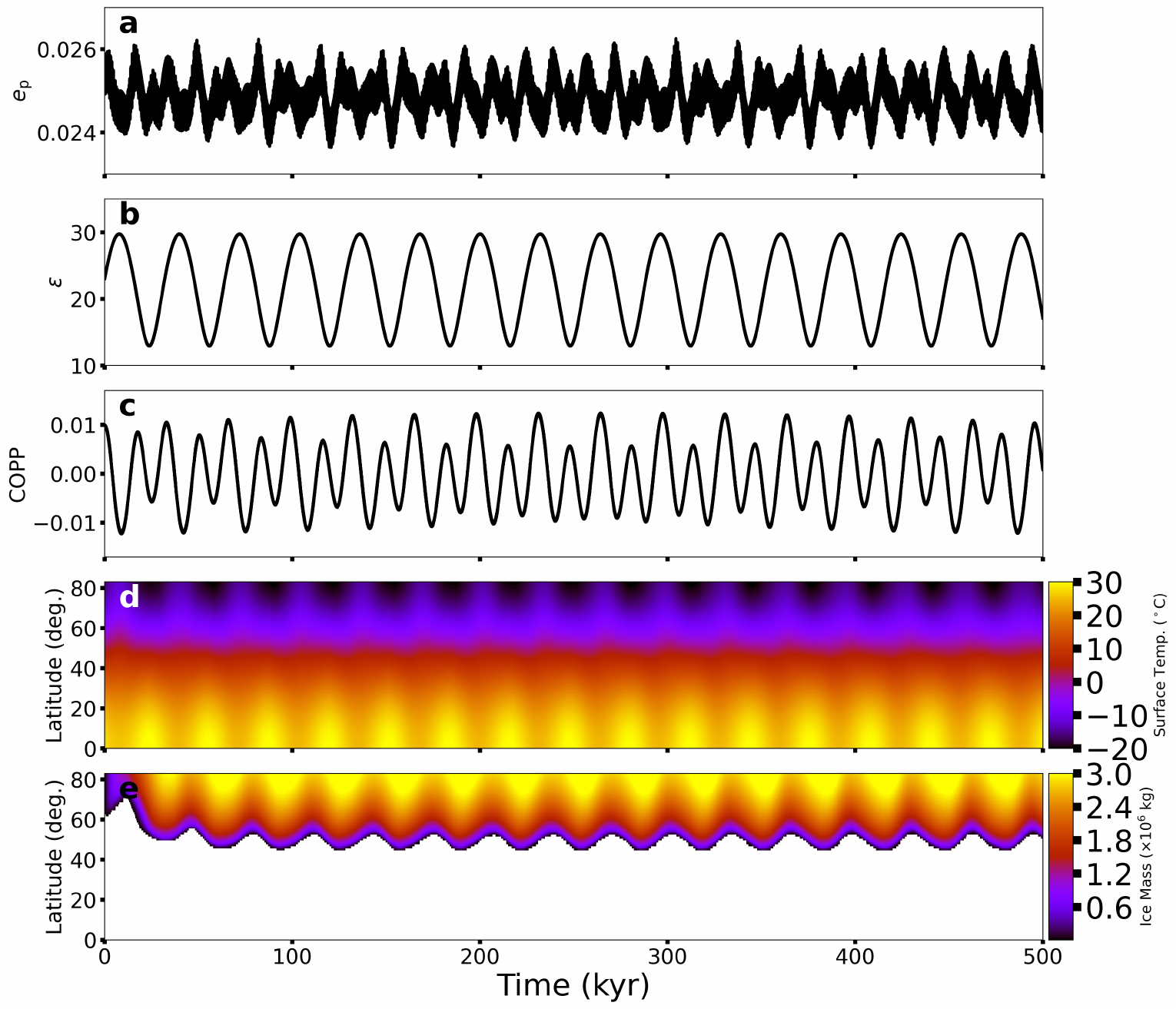}
    \caption{Similar to Fig. \ref{fig:EBM_ts_B_2}, but for an Earth-analog initially inclined 10$^\circ$ above the binary orbital plane. \label{fig:EBM_ts_B_10} }
\end{figure}

\begin{figure}
    \centering
    \includegraphics[width=\linewidth]{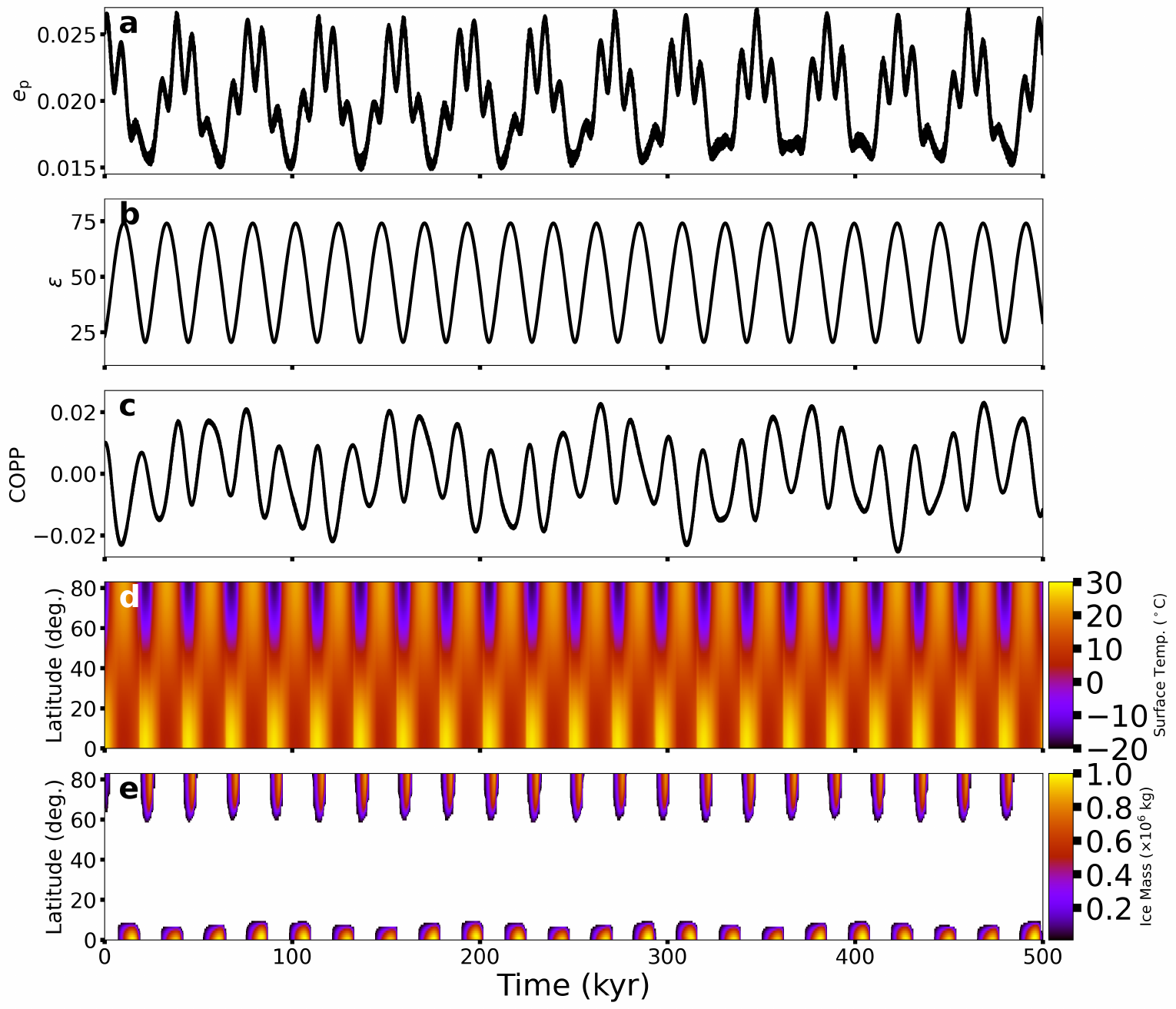}
    \caption{Similar to Fig. \ref{fig:EBM_ts_B_2}, but for an Earth-analog initially inclined 30$^\circ$ above the binary orbital plane. \label{fig:EBM_ts_B_30} }
\end{figure}

Figures \ref{fig:EBM_ts_B_2}--\ref{fig:EBM_ts_B_30} demonstrate the time evolution of the planetary eccentricity and obliquity, which affect the evolution of the {COPP} (Figs. \ref{fig:EBM_ts_B_2}c, \ref{fig:EBM_ts_B_10}c, and \ref{fig:EBM_ts_B_30}c),  surface temperature with latitude (Figs. \ref{fig:EBM_ts_B_2}d, \ref{fig:EBM_ts_B_10}d, and \ref{fig:EBM_ts_B_30}d), and the ice mass with latitude (Figs. \ref{fig:EBM_ts_B_2}e, \ref{fig:EBM_ts_B_10}e, and \ref{fig:EBM_ts_B_30}e) for each respective planetary inclinations (2$^\circ$, 10$^\circ$, and 30$^\circ$). Each of the cases presented (Figs. \ref{fig:EBM_ts_B_2}--\ref{fig:EBM_ts_B_30}) have identical spin precession constants (46\arcsec/yr), but have different orbital precession frequencies.  In Fig. \ref{fig:EBM_ts_B_2}c, the {COPP} is largely correlated with the $\sim$6$^\circ$ obliquity variation (Fig. \ref{fig:EBM_ts_B_2}b), which is imprinted in the variations of surface temperature (Fig. \ref{fig:EBM_ts_B_2}d) near the equator and the poles.  Fluctuations in the ice mass (Fig. \ref{fig:EBM_ts_B_2}e) exhibit Milankovitch cycles similar to Earth's, where the variation in the polar cap is due to a combination of precessions from the orbit and obliquity.

A planet inclined by 10$^\circ$ experiences larger obliquity variations ($\sim$18$^\circ$; Fig. \ref{fig:EBM_ts_B_10}b), but the {COPP} frequency increases due to the faster orbital precession (Fig. \ref{fig:EBM_ts_B_10}c).  Consequently, the planet has stronger fluctuations in the surface temperature (Fig. \ref{fig:EBM_ts_B_10}d) and ice mass (Fig. \ref{fig:EBM_ts_B_10}e).  The persistence of the polar caps is unaffected, but the extent and duration of interglacial periods at 50$^\circ$ latitude are increased due to the larger obliquity variation and faster orbital precession.  A planet inclined by 30$^\circ$ has obliquity variations of 50$^\circ$ from nodal precession, but the apsidal locking at the forced eccentricity is much weaker and allows for more substantial eccentricity variations (Figs. \ref{fig:EBM_ts_B_30}a and \ref{fig:EBM_ts_B_30}b).  The {COPP} is no longer only affected by the differing precession rates resulting in a more complex evolution (Fig. \ref{fig:EBM_ts_B_30}c).  The more dramatic obliquity variation results in periodic cooling and warming of the high latitudes corresponding to low and high points in the obliquity evolution (Fig. \ref{fig:EBM_ts_B_30}d).  Thus, there are polar caps for brief periods followed by much longer interglacial periods.  Since the obliquity rises above 55$^\circ$ \citep{Williams2003,Rose2017,Kilic2018}, ice {coverage} accumulates around the equator to form a so-called ``ice belt'' (Fig. \ref{fig:EBM_ts_B_30}e).  Similar to the polar caps, the ice belt is short lived and there is a brief ice free period between the two states.  The timescale to cycle through all three ice distribution states ($\sim$25,000 yr) and happens to be $\sim$1.5$\times$ the induced orbital precession from the binary companion ($\sim$17,000 yr).

\subsection{Obliquity and Global Surface Temperature Variation} \label{sec:aCenB_obl_var}
\begin{figure}
    \centering
    \includegraphics[width=\linewidth]{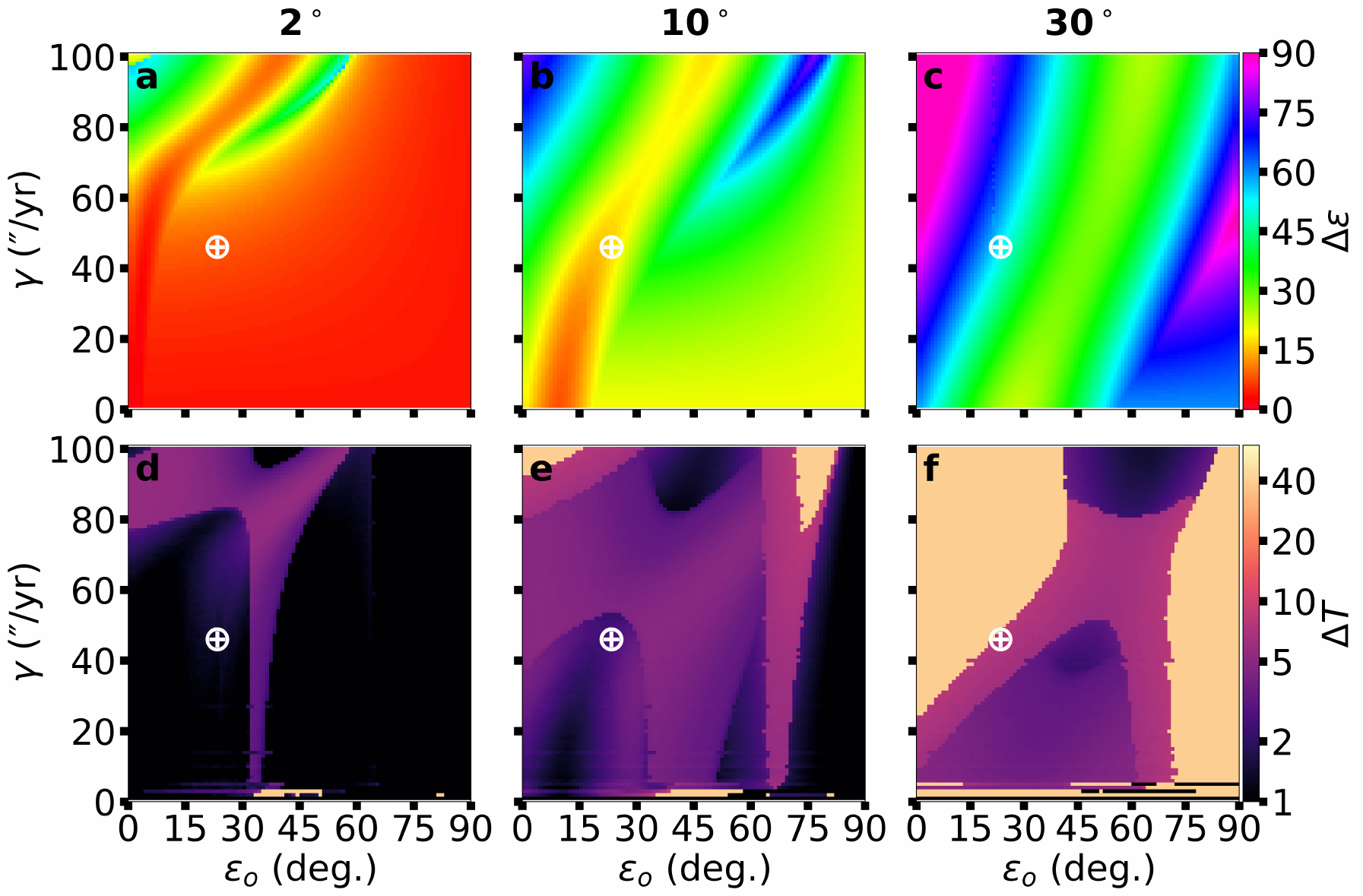}
    \caption{Obliquity and global surface temperature variation for an Earth-analog orbiting $\alpha$ Cen B over 400 kyr, where the planet's orbit is inclined by 2$^\circ$, 10$^\circ$, or 30$^\circ$ relative to the binary orbit.  The obliquity variation $\Delta \varepsilon$ (a--c) changes with the orbital inclination, with a valley of minimum variation resulting from orbital precession \citep{Quarles2019}.  The global surface temperature variation $\Delta T$ (d--f) largely correlates with the obliquity variation for a respective inclination.  The white $\oplus$ symbol designates conditions ($\varepsilon_o=23^\circ$, $\gamma = 46$\arcsec/yr, $P_{\rm rot}\sim24$ hr) for a modern, moonless Earth.  {Note that the color scales are different in this figure and Fig. \ref{fig:aCenB_ice_albedo}, when comparing with a similar figure for Star A (Fig. \ref{fig:aCenA_all}).}  \label{fig:aCenB_temp}}
\end{figure}

For an Earth-analog in $\alpha$ Cen A's HZ, the obliquity variations are relatively moderate ($\lesssim40^\circ$), even when considering a wide range of precession constants.  The HZ for $\alpha$ Cen B is more interesting because it allows for overlap in the secular orbital and spin precession frequencies so that larger obliquity variations are possible \citep{Quarles2019}.  We perform simulations for a broad range of precession constants and initial prograde obliquities following the same procedure discussed in Section \ref{sec:param_A}, but for three planetary inclinations (2$^\circ$, 10$^\circ$, and 30$^\circ$).  Figure \ref{fig:aCenB_temp} shows the obliquity variation ($\Delta\varepsilon$) and global surface temperature variation ($\Delta T$) using our numerical simulations over 500 kyr timescales, where the column labels denote the initial planetary inclination relative to the binary orbital plane.  The white $\oplus$ symbol designates initial parameters corresponding to a modern, moonless Earth-analog in terms of the precession constant (46\arcsec/yr) and initial obliquity (23$^\circ$).  Figure \ref{fig:aCenB_temp}a clearly shows the strip of low obliquity variation expected from strong spin-orbit coupling, which intrudes upon a broader region of increased obliquity variation at high spin precession constants (yellow-light blue) and these higher variations are due to a spin-orbit resonance with the binary.  Figure \ref{fig:aCenB_temp}b reproduces some of our previous results (see \cite{Quarles2019} for more details), while Fig. \ref{fig:aCenB_temp}c illustrates that moderate to large obliquity variations ($\Delta \varepsilon>15^\circ$) are ubiquitous for high inclination planets due to the nodal orbital precession.  Moreover, some combinations of spin precession and initial obliquity (magenta regions in Fig. \ref{fig:aCenB_temp}c) allow for the full range of prograde obliquity to be explored.

The obliquity variation in Figs. \ref{fig:aCenB_temp}a--\ref{fig:aCenB_temp}c presents one aspect, where our calculations using a climate model introduces another facet when considering the potential climate for an Earth-analog orbiting $\alpha$ Cen B.  The global surface temperature variation ($\Delta T$) in Fig. \ref{fig:aCenB_temp}d is limited to $\sim$6$^\circ$, where these variations largely occur when the obliquity variation is high.  However, larger $\Delta T$ values also occur within the strip for strong spin-orbit coupling; these variations extend to low spin precession constants for initial obliquity of $\sim$30--35$^\circ$.  Our investigation of the ice distributions for an Earth-analog orbiting $\alpha$ Cen A (see Section \ref{sec:aCenA_ice}) uncovered oscillations between states, which are the likely cause for the additional structure seen here.  

Figures \ref{fig:aCenB_temp}e and \ref{fig:aCenB_temp}f show that increased obliquity variation can lead to more extreme changes in the global surface temperature, where the tan regions signify simulations that begin with Earth-like global surface temperatures, but undergo climate feedbacks that force the planet into a snowball state (tan regions).  In particular, the regions for an Earth-analog to enter a snowball are especially large in Fig. \ref{fig:aCenB_temp}f, where a modern, moonless Earth-analog lies near a transition region of large temperature variation ($\sim$5$^\circ$; Fig. \ref{fig:EBM_ts_B_30}) and a snowball state.  Thus far, our discussion is limited to a moonless Earth-analog and adding a Luna-like moon (i.e., similar in mass and orbital separation to our Moon) would transport some the planet across various transition regions.  \cite{Quarles2019} showed the precession constant for an Earth-analog orbiting $\alpha$ Cen B with a Luna-like moon is $\sim$84\arcsec/yr, so such a moon would drastically reduce the potential for habitability of an Earth analog around $\alpha$ Cen B, in contrast to current conditions for the Earth-Moon system.

The tan strips for low precession constant ($\gamma\lesssim1$\arcsec/yr) could be spurious due to assumptions within \texttt{POISE} concerning the planetary rotation period becoming commensurable with the orbital period.  For an Earth-analog orbiting at the inner edge of $\alpha$ Cen B's HZ, a 2\arcsec/yr spin precession constant corresponds to a $\sim$24 day rotation period, which is an order of magnitude smaller than the orbital period and is consistent with the physical assumptions made by the developers of \texttt{POISE}.

\subsection{Ice Fraction and Distributions} \label{sec:aCenB_ice}
\begin{figure}
    \centering
    \includegraphics[width=\linewidth]{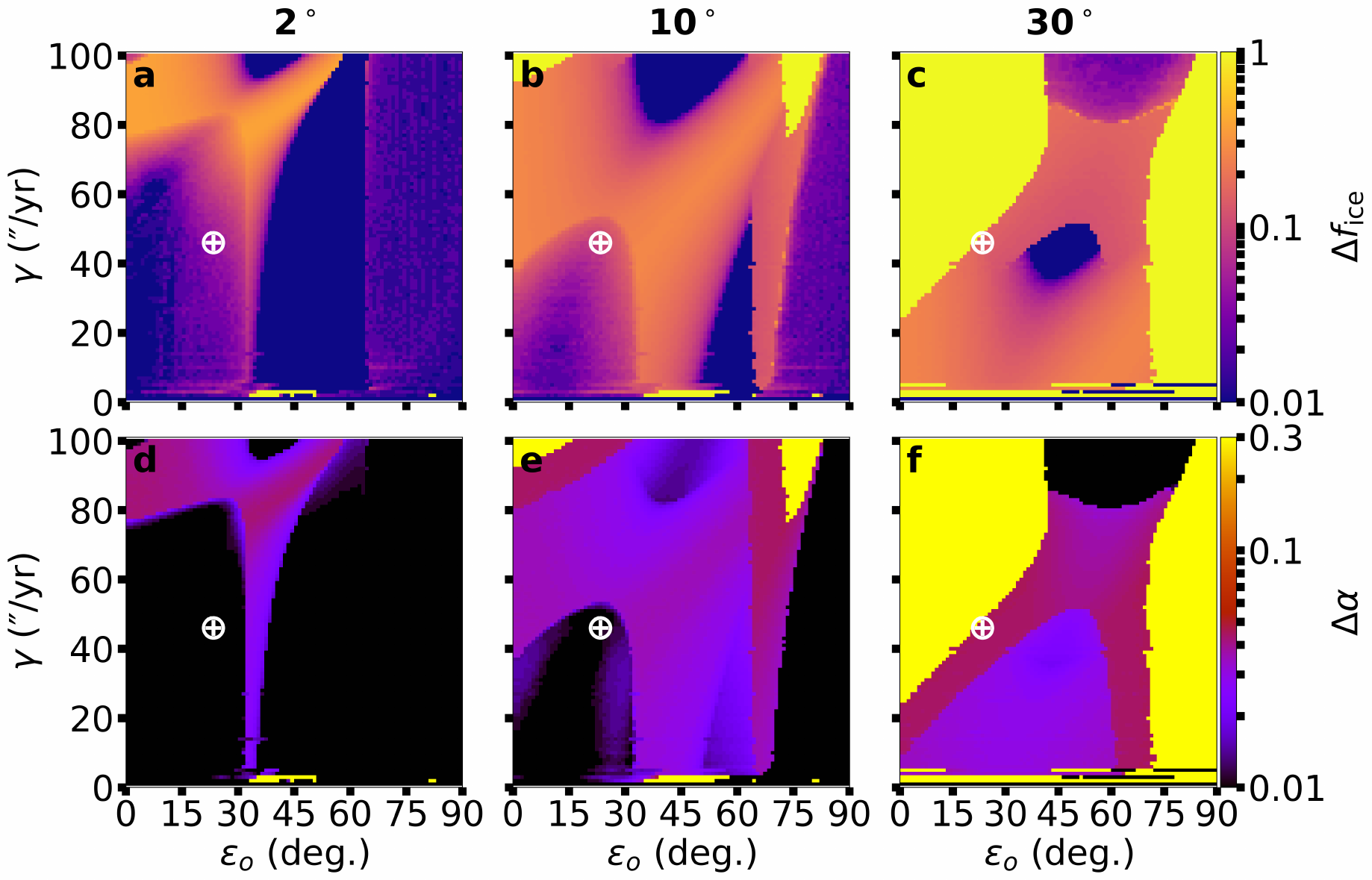}
    \caption{Similar to Fig. \ref{fig:aCenB_temp}, but for the global ice fraction $\Delta f_{\rm ice}$ (a--c) and albedo $\Delta \alpha$ (d--f) variations.  The white $\oplus$ symbol designates conditions ($\varepsilon_o=23^\circ$, $\gamma = 46$\arcsec/yr, $P_{\rm rot}\sim24$ hr) for a modern, moonless Earth.  \label{fig:aCenB_ice_albedo}}
\end{figure}

Moderate changes in the global surface temperature over time can correlate with variations in the global ice fraction (see Fig. \ref{fig:EBM_ts}).  Figure \ref{fig:aCenB_ice_albedo} broadly mimics the structures shown in the Figs. \ref{fig:aCenB_temp}d--\ref{fig:aCenB_temp}f for the respective planetary mutual inclination.  However, Figs. \ref{fig:aCenB_ice_albedo}a and  \ref{fig:aCenB_ice_albedo}b show more detail surrounding the main structures.  There are also particular regions where the global ice fraction $f_{\rm ice}$ changes very little ($\Delta f_{\rm ice}\lesssim 0.01$), in contrast to the yellow regions ($\Delta f_{\rm ice} = 1$ and $\Delta \alpha = 0.3$) that mark configurations that transition to snowball states.  The global albedo directly affects the calculation of the global surface temperature, and thus it is unsurprising that Figs. \ref{fig:aCenB_ice_albedo}d--\ref{fig:aCenB_ice_albedo}f are highly correlated with \ref{fig:aCenB_temp}d--\ref{fig:aCenB_temp}f.

\begin{figure}
    \centering
    \includegraphics[width=\linewidth]{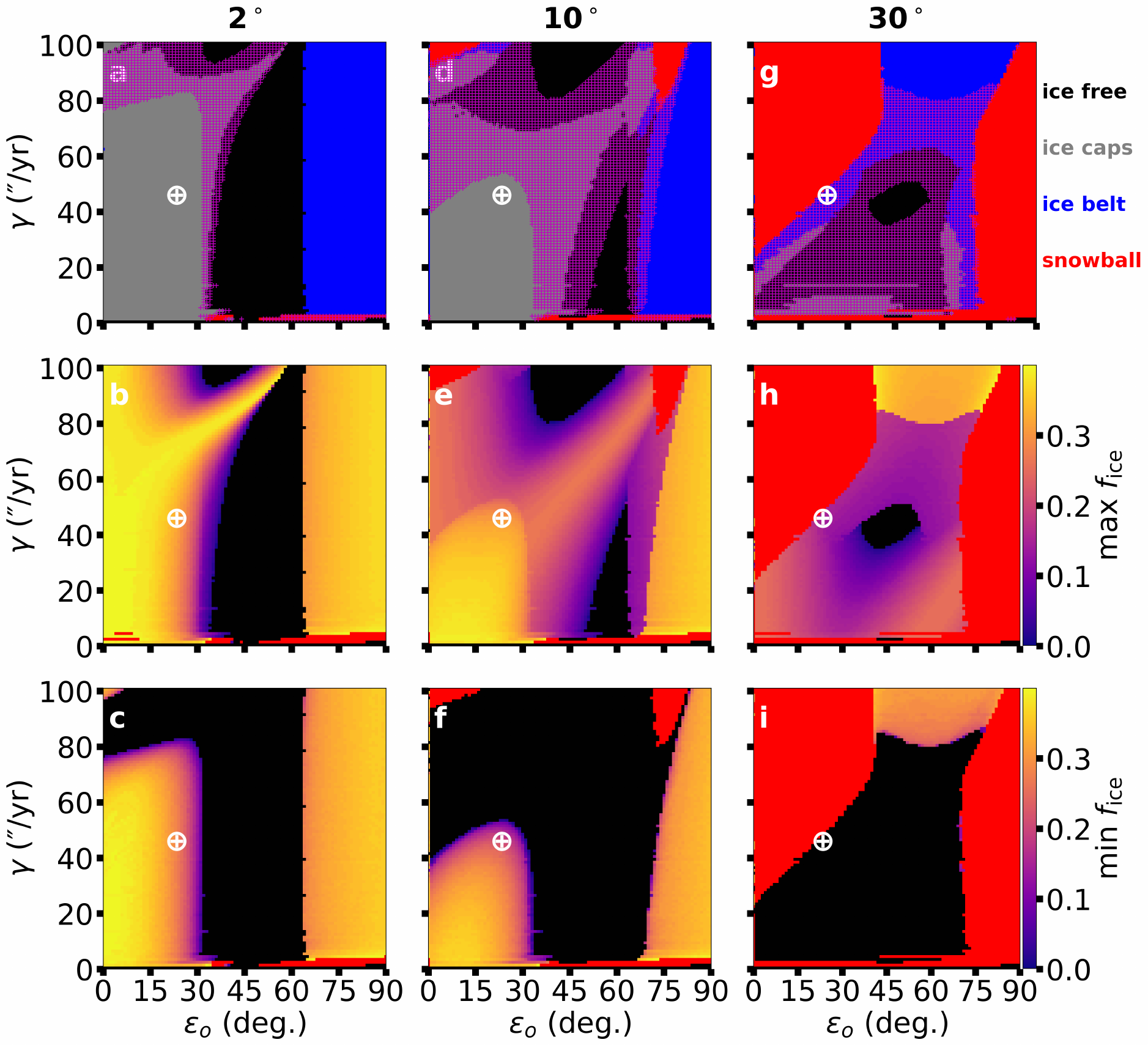}
    \caption{Similar to Fig. \ref{fig:aCenA_ice}, but examining differences in the most frequent ice distribution state (ice free, ice caps, ice belt, or snowball) as well as the maximum and minimum global ice fraction for a Earth-analog orbiting $\alpha$ Cen B inclined by $2^\circ$,  $10^\circ$, and  $30^\circ$ relative to the binary orbital plane.  The {magenta hatched region} plotted in (a) {represents the cases where} the planet oscillates between two or three ice distribution states. \label{fig:aCenB_ice} }
\end{figure}

Examining the variation in the global ice fraction uncovers the major structure across the full range of initial parameters ($\varepsilon_o$, $\gamma$) for the planetary spin states, but it leaves many questions.  Figure \ref{fig:aCenB_ice} shows the ice distribution categories, maximum global ice fraction, and minimum global ice fraction in a similar manner as Fig. \ref{fig:aCenA_ice}, but for an Earth-analog orbiting $\alpha$ Cen B and for our three planetary mutual inclinations by column.  There is also a white $\oplus$ symbol to mark initial conditions for a modern, moonless Earth-analog.  For nearly coplanar systems (Figs. \ref{fig:aCenB_ice}a--\ref{fig:aCenB_ice}c), we find a broad zone with ice caps for $\varepsilon_o \lesssim 30^\circ$, with ice free states for $30^\circ \lesssim \varepsilon_o \lesssim 60^\circ$, and with an ice belt for $60^\circ \lesssim \varepsilon_o \lesssim 90^\circ$.  Some slow rotators ($\gamma \lesssim 5$\arcsec/yr) can enter snowball states, but these conditions likely need more sophisticated models or at least an independent means of verification.  The panels illustrating the maximum and minimum $f_{\rm ice}$ are color-coded, where black and red corresponds to an exactly ice free or snowball state, respectively.  The bridge in Fig. \ref{fig:aCenB_ice}b for high $\gamma$ and moderate $\varepsilon_o$ forms due to the obliquity variation (see Fig. \ref{fig:aCenB_temp}a).  The common regions between Fig. \ref{fig:aCenB_ice}b and \ref{fig:aCenB_ice}c show which conditions can form stable ice caps (low $\varepsilon_o$) or an ice belt (high $\varepsilon_o$).  The difference regions between Fig. \ref{fig:aCenB_ice}b and \ref{fig:aCenB_ice}c show the limits of variation, where an Earth-analog oscillates between 2 or 3 states ({magenta hatching}).  Moreover, the regime that oscillates between ice free and ice caps highlights the respective structure observed in Figs. \ref{fig:aCenB_temp}d, \ref{fig:aCenB_ice_albedo}a, and \ref{fig:aCenB_ice_albedo}d.  Note that the background color signifies the most frequent state within our simulation, where the oscillation between ice distribution categories can go both ways.

Increasing the planet mutual inclination to 10$^\circ$ allows for larger obliquity variation, which enables a greater diversity of ice distribution states.  Figures \ref{fig:aCenB_ice}d--\ref{fig:aCenB_ice}f shows that the regime for primarily ice caps increases and an ice belt decreases.  The snowball regions correspond to  most conditions that produce the most  extreme obliquity variations (see Fig. \ref{fig:aCenB_temp}b).  Due to the overall higher obliquity variation, more of the parameter space can be ice free at times (Fig. \ref{fig:aCenB_ice}f), which suggests a greater potential for interglacial events.  The presence of a Luna-like moon could change an Earth-analog from a planet with perpetual ice caps to one with transient ice caps and ice free states being more frequent.  Our simulations show that snowball states become drastically more likely (due to higher obliquity variations; Fig. \ref{fig:aCenB_temp}c) for a highly inclined ($i_{\rm p} = 30^\circ$) Earth-analog, where a 24 hr rotator lies within a transition region (see Fig. \ref{fig:EBM_ts_B_30}e).  Perpetual ice caps are absent, while there is a much smaller regime for perpetually ice free states (black region near $\varepsilon_o\sim45^\circ$ in Fig. \ref{fig:aCenB_ice}g).  Figure \ref{fig:aCenB_ice}h shows that less of the planet's surface can support ice, unless the Earth-analog is in either a snowball state or a fast rotator ($\gamma\gtrsim80$\arcsec/yr).

\section{An Earth-analog orbiting the secondary star of  \texorpdfstring{$\alpha$}{alpha} Centauri-like binaries } \label{sec:starB}
The general population of Sun-like stellar binaries spans a large range in mass ratio, orbital period (or semimajor axis), and eccentricity \citep{Raghavan2010,Moe2017}. \cite{Quarles2019} explored the obliquity variations for an Earth-analog orbiting the more massive and luminous stellar component, star A.  In this work, we investigate the obliquity variations for an Earth-analog orbiting the secondary component, star B, and identify the possible consequences for the ice distributions including the Milankovitch cycles.  A range of binary semimajor axis and eccentricity values are used (see Section \ref{sec:orbit}), but we use stellar masses and luminosities that are identical to those for $\alpha$ Centauri AB.  The initial spin state begins with a 23$^\circ$ obliquity and 46\arcsec/yr spin precession constant, which is consistent with a modern, moonless Earth-analog.

\subsection{Factors Affecting Milankovitch Cycles} \label{sec:GenBin_Milan}
\begin{figure}
    \centering
    \includegraphics[width=\linewidth]{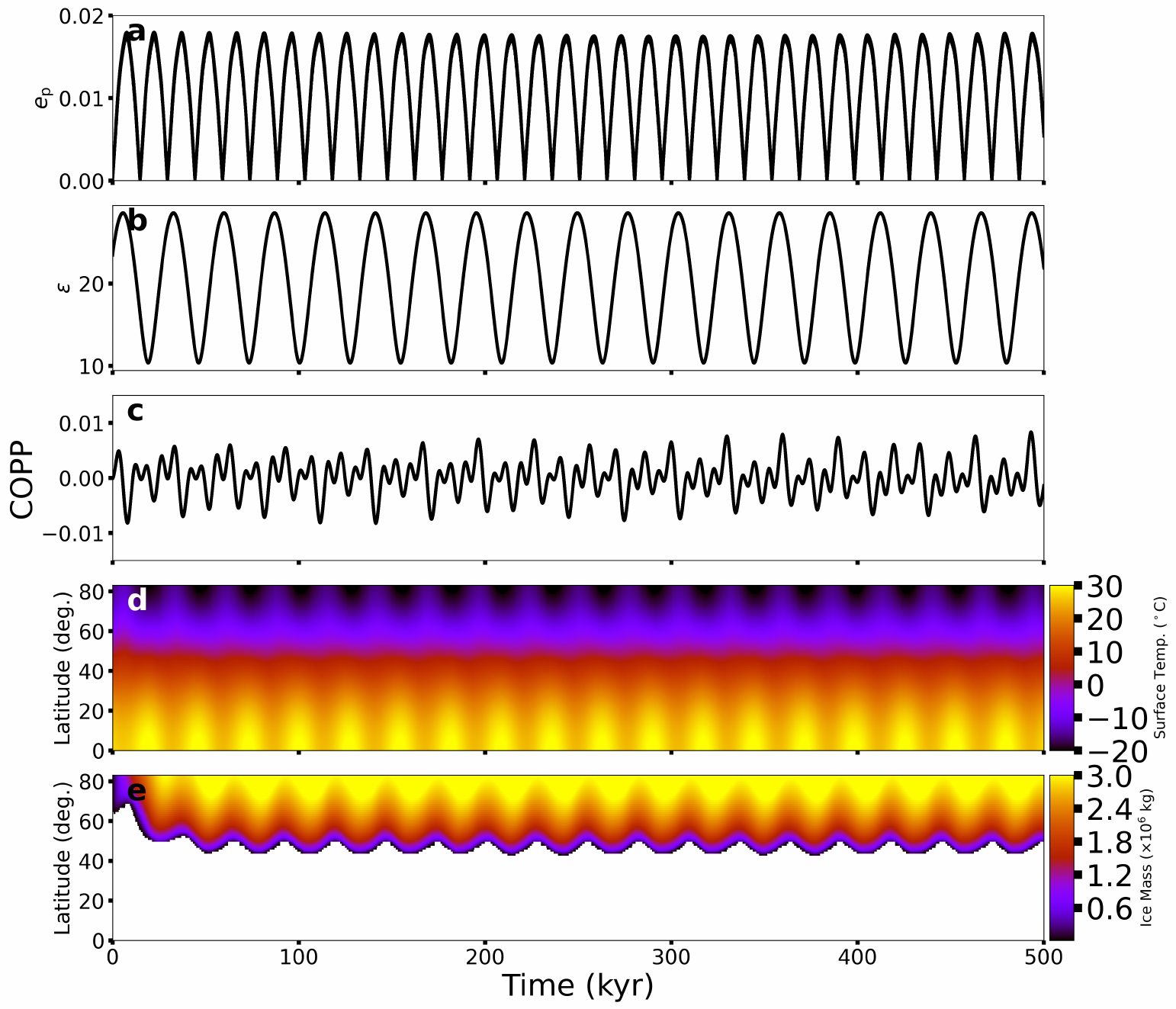}
    \caption{Evolution over 500 kyr of the (a) planetary eccentricity, (b) obliquity, (c) {COPP ($e_{\rm p}\sin{(\varepsilon)}\sin(\varpi_{\rm p} + \psi)$)}, (d) surface temperature, and (e) ice mass for an Earth-analog orbiting the secondary star in an $\alpha$ Centauri-like system.  The planet begins on an inclined, circular orbit that is tilted 10$^\circ$ relative to the binary orbital plane with a binary semimajor axis and eccentricity equal to 20 au and 0.2, respectively. The initial planetary spin state is analogous to a modern, moonless Earth ($\varepsilon_o=23^\circ$, $\gamma = 46$\arcsec/yr).   \label{fig:GenBin_ts_20} }
\end{figure}

\begin{figure}
    \centering
    \includegraphics[width=\linewidth]{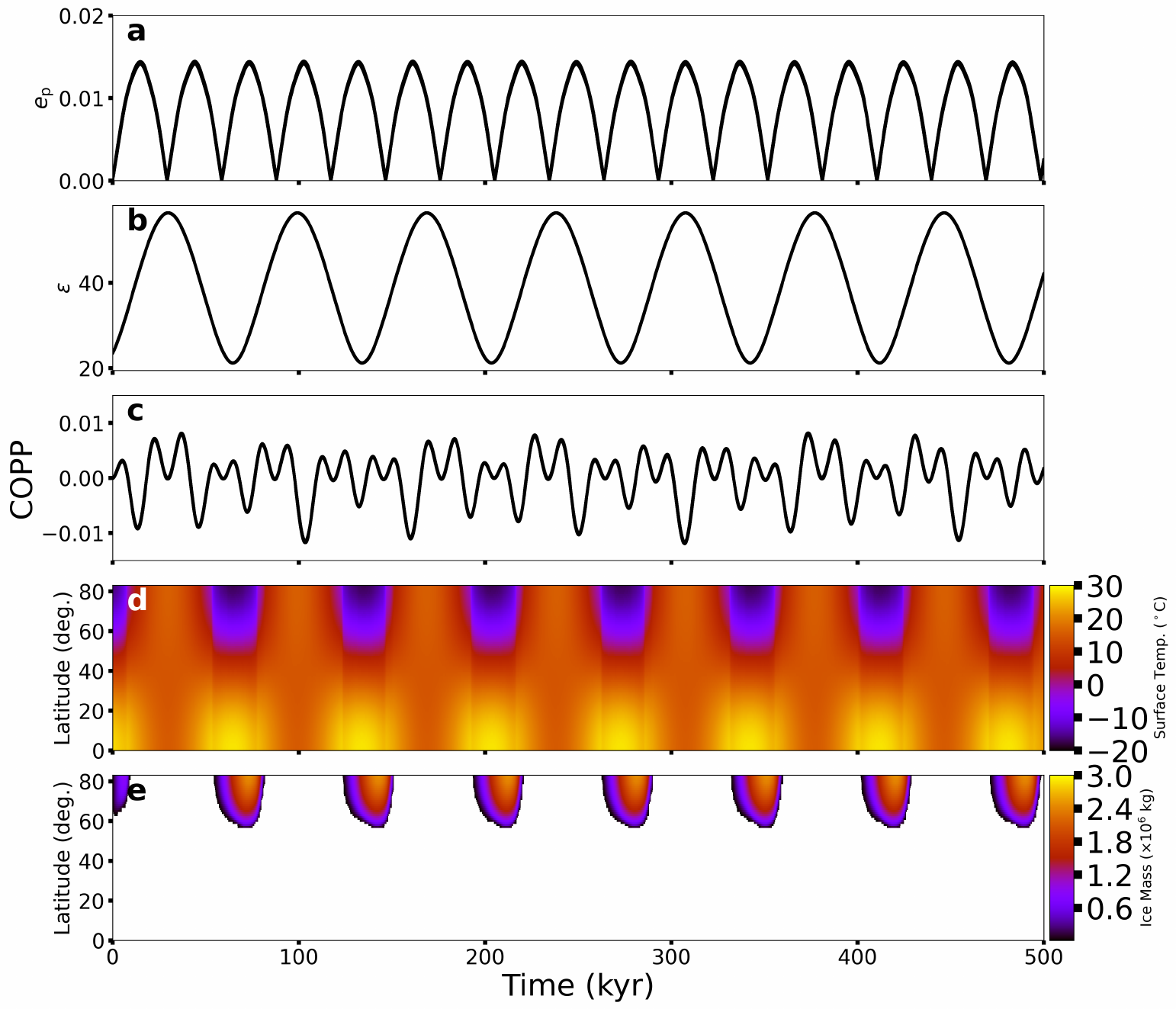}
    \caption{Similar to Fig. \ref{fig:GenBin_ts_20}, but the binary semimajor axis is 25 au.   \label{fig:GenBin_ts_25} }
\end{figure}

\begin{figure}
    \centering
    \includegraphics[width=\linewidth]{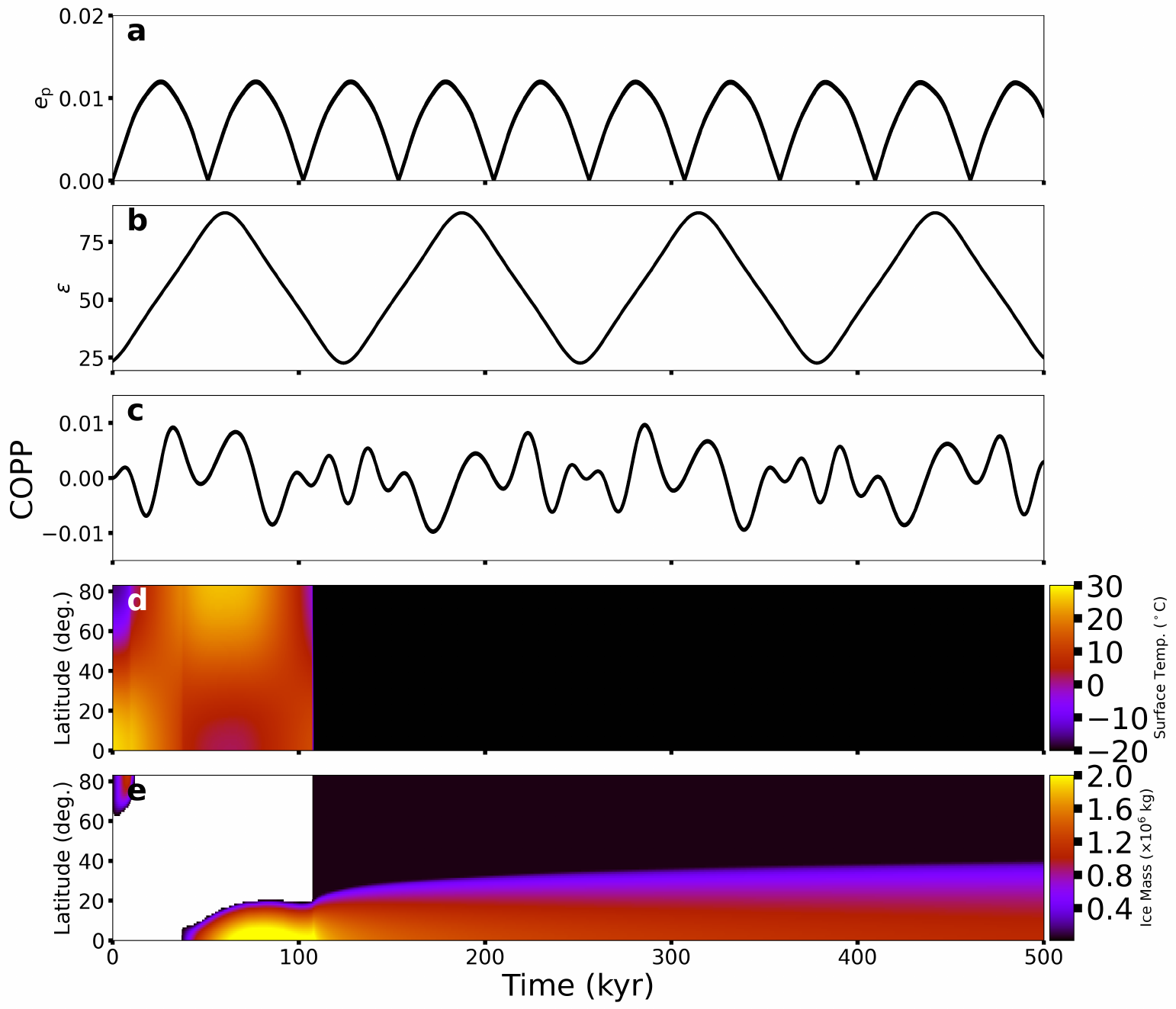}
    \caption{Similar to Fig. \ref{fig:GenBin_ts_20}, but the binary semimajor axis is 30 au.   \label{fig:GenBin_ts_30} }
\end{figure}

Milankovitch cycles for the Earth are shaped mainly by the evolution in its eccentricity and obliquity \citep{Milankovitch1969}, where the variations in these components are correlated with perturbations from neighboring terrestrial planets and Jupiter \citep{Laskar1993a,Deitrick2018}.  \cite{Deitrick2018b} showed that a large planetary eccentricity ($>0.1$) can trigger snowball states or even asymmetries in coverage of the ice caps.  For the case considered herein, there are no other planets but the stellar companion drives the orbital and spin evolution \citep{Andrade2016,Quarles2019}. Extreme obliquity variations can arise for the Earth-analog due to a spin-orbit resonance with stellar binary.  Figure \ref{fig:GenBin_ts_20}--\ref{fig:GenBin_ts_30} show the time evolution of the (a) planetary eccentricity, (b) obliquity, (c) {COPP}, (d) surface temperature, and (e) ice mass in a similar manner as in Sections \ref{sec:aCenA} and \ref{sec:aCenB}, but the binary semimajor axis is varied from 20, 25, and 30 au while the binary eccentricity begins at 0.2.  The Earth-analog begins on a circular orbit that is inclined 10$^\circ$ relative to the binary orbital plane.  The eccentricity variations for the Earth-analog remain quite low ($\leq 0.02$) and decrease substantially as the binary orbital semimajor axis increases, so the forced eccentricity plays a minor role in the climate evolution.

Figure \ref{fig:GenBin_ts_20} shows the planetary eccentricity changes with the secular forcing timescale ($\sim$15,000 yr) from the binary companion in Fig. \ref{fig:GenBin_ts_20}a, while obliquity variation timescale is a bit longer in Fig. \ref{fig:GenBin_ts_20}b.  These factors combine, along with their associated precession rates, to produce a more complex {COPP} in Fig. \ref{fig:GenBin_ts_20}c that drives variations in the surface temperature (Fig. \ref{fig:GenBin_ts_20}d) and ice mass (Fig. \ref{fig:GenBin_ts_20}).  In this simulation, the Earth-analog develops polar caps that are persistent and the ice latitude (i.e., border between white and colored cells) in Fig. \ref{fig:GenBin_ts_20}e correlates with the $\sim$18$^\circ$ obliquity variation (Fig. \ref{fig:GenBin_ts_20}b).

Increasing the binary semimajor axis to 25 au (Figure \ref{fig:GenBin_ts_25}) reduces the magnitude and frequency of the eccentricity variations (Fig. \ref{fig:GenBin_ts_25}a).  Consequently, the secular precession frequency begins to approach the spin precession frequency, which causes an increase in the obliquity variation ($\Delta\varepsilon\sim$30$^\circ$; Fig. \ref{fig:GenBin_ts_25}b).  The climate precession frequency decreases (Fig. \ref{fig:GenBin_ts_25}c), although a similar structure exists that is due to the linear combination of frequencies.  A longer climate precession timescale and larger obliquity variation permits the Earth-analog to oscillate between a warm and temperate climate (Fig. \ref{fig:GenBin_ts_25}d).  The warmer climate occurs when the obliquity reaches 40$^\circ$--60$^\circ$ and the polar caps completely disappear.  In this state the equatorial latitudes are cooler, but the poles are significantly warmer.  Once the obliquity decreases below 40$^\circ$, the temperature at high latitudes drops low enough to allow for polar caps to regrow.  Figure \ref{fig:GenBin_ts_25}e shows the cycle between interglacial (ice free) states and transient polar caps.

The secular forcing frequency due to the binary companion overlaps with the spin precession frequency at $\sim$30 au, which induces the largest obliquity variations.  The forced eccentricity oscillations (Fig. \ref{fig:GenBin_ts_30}a) with a 30 au binary separation are small and the $\sim$60$^\circ$ obliquity variation (Fig. \ref{fig:GenBin_ts_30}b) largely determines how climate evolves on the Earth-analog.  The {COPP} undergoes cycles every $\sim$125 kyr (Fig. \ref{fig:GenBin_ts_30}c), but the diminished magnitude of the maximum eccentricity reduces its influence upon the Milankovitch cycles.  Figure \ref{fig:GenBin_ts_30}b shows the planetary obliquity rises above 55$^\circ$ for a significant portion of its cycle, which leads to the formation of an ice belt instead of an ice cap \citep{Kilic2018}.  Figure \ref{fig:GenBin_ts_30}d illustrates this transition, where the equatorial latitudes have a much higher surface temperature for low obliquity and drop to freezing temperatures once the obliquity increases enough.  As a result, the Earth-analog develops small ice caps that eventually disappear, which is followed by a significant ice belt (Fig. \ref{fig:GenBin_ts_30}e).  As the obliquity decreases,  we might expect the ice belt to fully ablate followed by the growth of an ice cap (see Fig. \ref{fig:EBM_ts_B_30}e).  However, this is not the case because the ablation rate of the ice belt is too slow (i.e., thermal inertia of the ice belt is high).  Under these conditions, the ice cap begins to grow before the ice belt disappears and an ice-albedo feedback prevents the return to an ice cap dominated state (i.e., a majority of the surface is ice covered, which stops the heat flow across latitudes).  Instead, the Earth-analog enters a snowball state, which persists indefinitely in our simple model.  \cite{Deitrick2018b} showed a similar ice instability for dynamically hot planetary systems orbiting a single G dwarf.  We illustrate one pathway to a snowball state, but this is not destiny for all planets because of the limitations of our climate model that are calibrated for Earth-like conditions and other features (e.g., land/ocean fraction, CO$_2$ warming, land distribution) could change the heat flow across latitudes.

\subsection{Obliquity and Global Surface Temperature Variation}
\begin{figure}
    \centering
    \includegraphics[width=\linewidth]{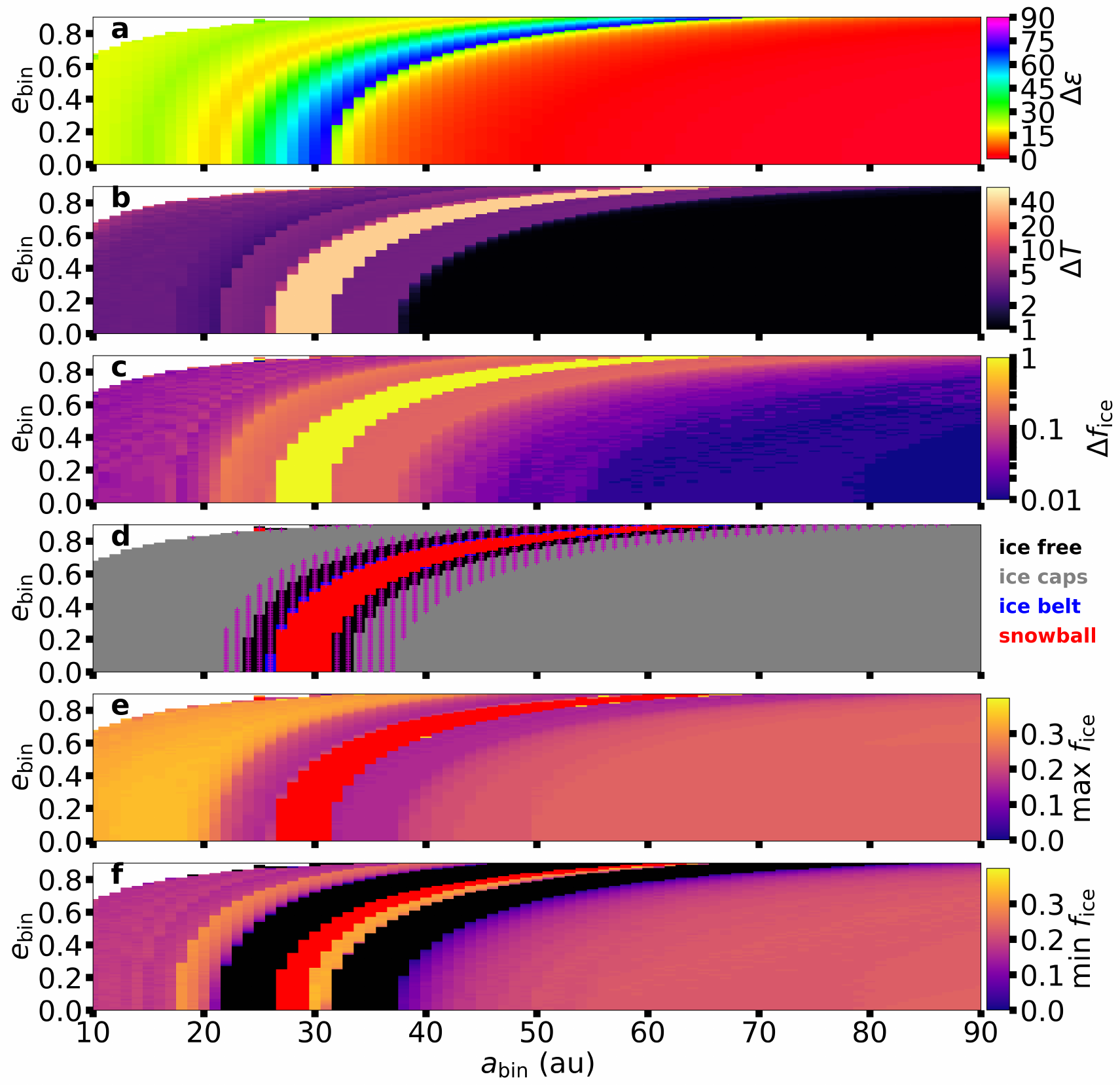}
    \caption{Results of simulations for an Earth-analog orbiting the secondary star of an $\alpha$ Centauri-like binary, where the binary semimajor axis $a_{\rm bin}$ and eccentricity $e_{\rm bin}$ are varied.  The (a) obliquity, (b) global surface temperature, and (c) global ice fraction variations illustrate general structures that impact the potential habitability of the planet.  The (d) ice distribution state, (e) maximum ice fraction, and (f) minimum ice fraction show how the binary orbit affects the potential Milankovitch cycles.  The white cells (top left) indicate orbital parameters, where the planetary orbit would not be stable \citep{Quarles2020b}.  The {magenta hatched region} plotted in (d) {represents the cases where} the planet oscillates between two or three ice distribution states.   \label{fig:GenBin}}
\end{figure}

For an Earth-analog within a binary star system, significant obliquity variations exist when the secular forcing frequency from the binary is larger than the spin precession frequency (i.e., $g_{\rm s}\gtrsim \gamma$).  \cite{Quarles2019} showed this feature for an Earth-analog orbiting the more massive star; herein we perform a similar investigation of an Earth-analog orbiting the less massive star in Figure \ref{fig:GenBin}a.  The white cells in each panel of Fig. \ref{fig:GenBin} are not evaluated due to orbital instabilities within the host star's HZ \citep{Quarles2020b}.  Closely separated binaries ($a_{\rm bin}$=10--20 au) induce moderate obliquity variations ($\Delta\varepsilon\sim20^\circ-40^\circ$), while there is a yellow strip of lower obliquity variation ($\Delta\varepsilon\sim15^\circ$) beginning at 20 au.  This strip occurs when the planetary spin axis precession more closely follows the orbital precession (i.e., stronger spin-orbit coupling).  The curvature of the strip stems from how the secular forcing frequency $g_{\rm s}$ depends on the binary semimajor axis, $a_{\rm bin}$, and eccentricity, $e_{\rm bin}$.  For $a_{\rm bin}=22-27$ au, the obliquity variation increases to $\sim$45$^\circ$, and the highest variation occurs for $a_{\rm bin}=28-31$ au due to a spin-orbit resonance.  For most configurations with $a_{\rm bin}>40$ au, the spin-orbit coupling is stronger, which limits the obliquity variation to a few degrees.

The global temperature (Fig. \ref{fig:GenBin}b) and ice fraction variation (Fig. \ref{fig:GenBin}c) largely correlate with the obliquity variations in Fig. \ref{fig:GenBin}a.  The global temperature can oscillate by  5$^\circ$C--10$^\circ$C (purple), decrease dramatically by 40$^\circ$ (tan), or remain nearly constant (black) in response to the changes in obliquity.  The large obliquity variations due to the spin orbit resonance force the large global temperature change through ice-albedo feedbacks (see the Section. \ref{sec:GenBin_Milan} and Fig. \ref{fig:GenBin_ts_30}).  We find the global temperature variations broadly correlate with changes in the global ice fraction (Fig. \ref{fig:GenBin}c).

\subsection{Ice Fraction and Distributions}
Following Sections \ref{sec:aCenA_ice} and \ref{sec:aCenB_ice}, we examine the ice distribution within four categories (ice free, ice caps, ice belt, and snowball), but through the parameter space of varying the binary orbit.  Figure \ref{fig:GenBin}d shows the most common state for a modern, moonless Earth-analog is having perpetual ice caps (gray) with a varying degree of Milankovitch cycles (i.e., periodic growth/retreat at the ice latitudes), while the least common is having an ice belt.  This outcome could be flipped if a sufficiently high initial obliquity was used.  An Earth-analog enters a snowball (red) state when the obliquity variation is $\gtrsim$50$^\circ$, which occurs when the obliquity is $>55^\circ$ for a significant portion of the obliquity variation cycle.  Ice free states (black) are possible for moderate obliquity variation, when the obliquity is between 40$^\circ$--55$^\circ$ for a significant portion of the obliquity variation cycle.  Surrounding the region for spin-orbit resonance, an Earth-analog can transition between states for ice free, ice caps, and an ice belt as designated by the {magenta symbols}.

Figures \ref{fig:GenBin}e and \ref{fig:GenBin}f show the extremes in the global ice fraction $f_{\rm ice}$.  When the orbital perturbations on the Earth-analog are greatest ($a_{\rm bin}<20$ au), the global ice fraction oscillates between 0.15--0.38.  For lower obliquity variations ($\Delta\varepsilon\sim15^\circ$), the global ice fraction varies slightly (see Fig. \ref{fig:GenBin_ts_20}e), and ice caps cover about 1/3 of the total surface.  Comparing the 20--40 au region of Figs. \ref{fig:GenBin}d--\ref{fig:GenBin}f, the extent of the most common ice distributions can be deduced, where the largest extent of ice covers 1/5 of the total surface and ice free states are possible.  When $a_{\rm bin}>40$ au, ice caps dominate with little variation in their extent ($f_{\rm ice}=0.1-0.2$).  Currently, the global ice fraction on Earth is $\sim$0.1, but this fraction was larger in the recent geologic past (10 kyr ago) during an ice age with a northern ice cap that extended from $90^\circ$N--45$^\circ$N to create the Great Lakes.

% \section{Possible Observational Signatures through Albedo Variations} \label{sec:obs}

\section{Conclusions} \label{sec:conc}
The potential habitability of an Earth-analog is defined by the planet's ability to host liquid water upon its surface, and orbital perturbations can potentially alter a planet's habitability through Milankovitch cycles (i.e., growth/retreat of ice).  We investigate the obliquity evolution of a circumstellar Earth-analog within $\alpha$ Centauri-like binaries near the inner edge of the respective host star's HZ, where a variable obliquity can induce changes in the global ice fraction and thereby modify the global surface temperature.  Our simulations explore a wide range of spin states varying the initial obliquity, $\varepsilon_o$, and spin precession constant, $\gamma$.  We identify initial spin states for an Earth-analog that allow it to sustain persistent ice at either high latitudes (ice caps) or near the equator (ice belt;\cite{William2003,Rose2017,Kilic2018}) using a one-dimensional EBM called \texttt{POISE} within \texttt{VPLanet} \citep{Deitrick2018b,Barnes2019}.  Some initial spin states remain ice free or evolve into snowball states, where the latter could dramatically limit a planet's habitability.  {Note that our usage of \emph{ice cap} is different than the Earth science community and refers to ice coverage beginning at a pole and extending towards the equator.}

The obliquity variation ($\Delta\varepsilon$) is fairly mild for an Earth-analog inclined by 10$^\circ$ orbiting within the HZ of $\alpha$ Cen A (see Fig. \ref{fig:aCenA_all}) and is largely driven by the nodal precession induced by the stellar companion ($\Delta\varepsilon\sim 2i_{\rm p}$).  The maximum obliquity variation is $\sim$30$^\circ$ over 20 kyr timescale, which limits changes in the globally averaged ice fraction ($\Delta f_{\rm ice} \lesssim 0.2)$ and surface temperature ($\Delta T < 6^\circ$C).  Since the changes in ice fraction are small (mostly at border regions of ice caps), the resulting changes in albedo are also small ($<3\%$).  An Earth-analog can have persistent ice caps for low initial obliquity ($\lesssim30^\circ$) or can be completely ice free for high initial obliquity ($>60^\circ$).  Between these extremes, the global ice distribution can oscillate from ice free to ice caps (or vice versa) depending on the planet's initial obliquity and rotation rate (through the spin precession constant $\gamma$).  The Milankovitch cycles can be extreme for high $\gamma$ with growth/retreat of polar caps by $\sim20^\circ$ of latitude.

For an Earth-analog orbiting $\alpha$ Cen B, we find a more dynamic range of obliquity variation that depends on the initial inclination (see Fig. \ref{fig:aCenB_temp}).  The obliquity evolution is driven by nodal precession in broad regions, but these are interrupted by valleys of low variation due to strong spin-orbit coupling \citep{Quarles2019} and peaks of high obliquity variation from a spin-orbit resonance with the host binary. We investigate three orbital inclinations ($2^\circ$, $10^\circ$ and $30^\circ$). A nearly coplanar (2$^\circ$) Earth-analog orbiting $\alpha$ Cen B evolves in a similar manner to one orbiting $\alpha$ Cen A, but high initial obliquity ($>65^\circ$) states produce an ice belt instead of ice caps (Fig. \ref{fig:aCenB_ice}).  An Earth-analog inclined by 10$^\circ$ can be in a snowball or ice free state within two pockets within the parameter space.  Many initial spin states allow for the Earth-analog to transition periodically between ice free, ice caps, and an ice belt.  A highly inclined ($30^\circ$) Earth-analog can enter a snowball state for a much broader set of initial parameters.  Due to the higher overall obliquity variation ($\Delta\varepsilon\sim60^\circ$), the region dominated by an ice belt moves to lower initial obliquity.  Initial spin states that permit persistent ice caps become quite rare with a higher orbital inclination, and periodic transitions between ice distribution states becomes common (except when a snowball is formed). The Milankovitch cycles for a low inclination, moonless Earth-analog (24 hr rotator) are not drastic, but the cycles for an Earth-analog with a faster rotation or a higher initial obliquity allow transitions between states.  

\cite{Quarles2019} explored the effect of varying binary orbits on the obliquity variation using an Earth-analog inclined by $10^\circ$ orbiting the more massive star of an $\alpha$ Centauri-like binary, where we perform a similar investigation but the planet orbits the less massive star (Fig. \ref{fig:GenBin}).  We find similar trends, where the variations are shifted towards smaller binary semimajor axis $a_{\rm bin}$ due to a difference in the secular forcing frequency \citep{Andrade2016,Quarles2019}.  A majority of parameters for the binary orbit permit persistent ice caps with mild Milankovitch cycles, but this is partly due to our choice of initial obliquity, and an ice belt could become the most common if our simulations started the Earth-analog with a higher obliquity.  For $a_{\rm bin}$$\sim$$22-36$ au, Milankovitch cycles can be dramatic due to switching between states.  At the location of the spin-orbit resonance, the Earth-analog enters a snowball state because ice caps form before the previous phase of an ice belt fully ablates.  With significant ice {coverage} at both the equator and poles, the ice-albedo feedback quickly drives the planet into a snowball state.  This is consistent with the snowball instability discussed by \cite{Deitrick2018b} for an Earth-analog in a dynamically hot multiple planet system orbiting a Sun-like star.

\cite{Forgan2016} used a one-dimensional EBM similar to \texttt{POISE}, where the main differences are the heat capacity $C$ is parameterized instead of the albedo $\alpha$ and the orbital/spin evolution is more self-consistently coupled with the climate model.  One of the main conclusions of both their work and ours is that Milankovitch cycles for planets in binaries are influenced by the secular forcing of the binary, where \cite{Forgan2016} deduced this through a frequency analysis of the global surface temperature evolution.  This work goes beyond this prior analysis in many ways, specifically in the growth/retreat of ice and global distribution of ice.  Moreover, we examine a larger variation in the initial planetary spin state, planetary orbit, and binary orbit, where variations in these parameters can dramatically influence the Milankovitch cycles of the Earth-analog.

As the planetary orbital inclination relative to the binary orbit increases, so does the planet's obliquity variation due to nodal precession.  This can lead to more persistent ice caps, an ice belt, or even snowball states, but the first two of these do not necessarily diminish the habitability of the Earth-analog.  \cite{Ramirez2018} and \cite{Paradise2019} showed for both water worlds and for an Earth-like land/ocean fraction, respectively, that an increased CO$_2$ partial pressure or weathering can allow for some surfaces to be locally ice free.  \cite{Haqq-Misra2019} showed for circumbinary planets the latitudinal surface temperature can increase with different assumptions on the assumed planet topography and land/ocean fraction, where we expect similar effects are possible for eccentric circumstellar planets in binaries.  \texttt{POISE} has additional options to set the CO$_2$ partial pressure, change the constants for the outgoing longwave radiation, and mimic Hadley heat diffusion, where varying any one of these could be an interesting avenue for future work.  \texttt{VPLanet} has an open-source framework, allowing developers to introduce more sophisticated components to \texttt{POISE} or even new modules that go beyond the built-in assumptions \citep{Barnes2019}.

Identifying the existence of ice on exoplanets is a future endeavor for next generation telescopes that have yet to see first light (e.g., LUVOIR \citep{TheLUVOIRTeam2019} or HabEx \citep{Gaudi2020}), where albedo variations with respect to the planetary rotation and orbit could provide some of the necessary initial parameters \citep{Schwartz2016}.  Further atmospheric characterization with \textit{JWST} \citep{Gardner2006} would likely be necessary in trying to distinguish between the reflectivity of ice or clouds.  For $\alpha$ Centauri AB the initial step of planet detection has been elusive, but \cite{Beichman2020} proposed a path forward using \textit{JWST} and others in recent years are attempting to detect planets orbiting either star using data taken from ground-based facilities \citep{Trigilio2018,Zhao2018,Kasper2019,Wagner2021,Akeson2021}.

\section*{Acknowledgements}
This research was supported in part through research cyberinfrastructure resources and services provided by the Partnership for an Advanced Computing Environment (PACE) at the Georgia Institute of Technology. {The authors thank Rory Barnes for his guidance in using and modifying \texttt{VPLanet} as well as constructive comments that enabled us to improve the quality and clarity of the manuscript}.

%%%%%%%%%%%%%%%%%%%%%%%%%%%%%%%%%%%%%%%%%%%%%%%%%%
\section*{Data Availability}
Processed data and python scripts to reproduce the figures are available through the GitHub repository: \href{https://github.com/saturnaxis/Ice-ages-in-AlphaCen}{saturnaxis/Ice-ages-in-AlphaCen}.  The raw data from \texttt{VPLanet} underlying this article will be shared on reasonable request to the corresponding author.



\bibliographystyle{mnras}
\bibliography{sample63.bib} % if your bibtex file is called example.bib

%\begin{thebibliography}{99}
%\end{thebibliography}

%%%%%%%%%%%%%%%%%%%%%%%%%%%%%%%%%%%%%%%%%%%%%%%%%%

% Don't change these lines
\bsp	% typesetting comment
\label{lastpage}
\end{document}